\documentclass[preprint,superscriptaddress,showpacs]{revtex4}
\usepackage{graphicx}
\usepackage{amssymb}
\newcommand{\be}{\begin{eqnarray}}
\newcommand{\ee}{\end{eqnarray}}
\newcommand{\eq}{\begin{eqnarray}}
\newcommand{\en}{\end{eqnarray}}

\newcommand{\bfk}{{\bf k}_{\perp}}

\begin{document}

\title{Gravitational form factors and transverse spin sum rule in a light front quark-diquark model in AdS/QCD }
\author{Dipankar Chakrabarti}

\author{Chandan Mondal}
\affiliation{ Department of Physics, Indian Institute of Technology Kanpur, Kanpur 208016, India}

\author{Asmita Mukherjee}
\affiliation{ Department of Physics, Indian Institute of Technology Bombay, Powai, Mumbai 400076, India}

\date{\today}

\begin{abstract}
The gravitational form factors are related to the matrix elements of the energy-momentum tensor $T^{\mu\nu}$. Using the light front wave functions of the scalar quark-diquark model for nucleon predicted by the soft-wall AdS/QCD, we calculate the flavor dependent $A(Q^2)$, $B(Q^2)$ and $\bar{C}(Q^2)$ form factors. We also present all the matrix element of the energy-momentum tensor in a transversely polarized state. Further, we evaluate the matrix element of Pauli-Lubanski operator in this model and show that 
the intrinsic spin sum rule involves  the form factor $\bar{C}$. The longitudinal momentum densities in the transverse impact parameter space   are also discussed for both unpolarized and transversely polarized nucleons.

\end{abstract}
,

\maketitle

\section{Introduction}

Understanding the spin structure of the proton, which means how the proton
spin (1/2) is distributed among its constituent quarks and gluons is one of
the fundamental problems in hadron physics. Most of the studies, both
theoretical and experimental, mainly aim at the longitudinal spin or
helicity. Understanding the transverse spin and transverse angular
momentum of the proton is a much more involved problem. The complications
associated with the transverse angular momentum are best understood in
light-front framework, in which one gets an intuitive picture of deep
inelastic electron-proton scattering processes. The longitudinal angular
momentum operator is kinematical on the light-front, whereas the transverse
angular momentum and rotation operators are dynamical. This implies that the
partonic structure of the transverse spin is different from that of the
longitudinal spin structure of the proton. Recently, several sum rules have
been proposed in the literature about transverse spin. In \cite{hari2,hari3} a sum rule was
derived in terms of the intrinsic transverse spin operator on the
light-front. Unlike the transverse rotation operator or Pauli-Lubanski
operator, matrix elements of the intrinsic spin operator are frame
independent. This sum rule was explicitly verified in perturbation theory
for a dressed quark at one loop. Another sum rule was proposed in
\cite{leader3}
involving the transversity distribution. A new transverse polarization sum
rule was proposed in \cite{ji12,ji3,JXY} which was interpreted at the partonic level. This
was partially motivated by \cite{burkardt}  where a relation between the expectation
values of equal time transverse rotation operators and the gravitational
form factors is derived using delocalized states in the rest
frame of the nucleon. Authors of \cite{ji12,ji3} analyzed the matrix elements of the
transverse component of the Pauli-Lubanski operator for a transversely
polarized state and related it to the gravitational form factors $A(0)$ and
$B(0)$. In \cite{Leader1,HKMR,hari} it was pointed out that the above result is frame dependent.
In fact the only frame independent result is obtained in terms of the
intrinsic spin operators on the light-front; the corresponding relation not
only involves $A(0)$ and $B(0)$ but also the higher twist term $\bar C(0)$,
and the contribution from $\bar C(0)$ is not suppressed. In this work, we
verify the statements made in \cite{hari} in a model calculation.

Here, we evaluate the  GFFs for a transversely polarized proton from the energy momentum tensor and verify the sum rule for the transverse spin in a light front quark-diquark model. For this work, we take a phenomenological light front quark-diquark model recently proposed by Gutsche {\it et al.} \cite{Gutsche} with the corrected parameters given in Ref.\cite{MC}. In this model, the diquark is considered to be scalar(i.e. scalar diquark model) and the light front wave functions for the proton are constructed from the wave functions obtained in light front AdS/QCD correspondence\cite{BT}.  The parameters in this model are fixed by fitting to the electromagnetic form factors of the nucleons.  
Using the overlap formalism of light front wave functions,
we  calculate the GFFs from the energy momentum tensor ($T^{\mu\nu}$) for a transversely polarized proton. The intrinsic spin  operators  which can be derived from the transverse components of  the Pauli-Lubansky operator are shown to satisfy the sum rule consistent with \cite{hari}.

 The Fourier transform of the gravitational form factor  in the impact parameter space has interesting interpretations\cite{ selyugin,abidin08}. The Fourier transform of the form factor $A(Q^2)$ gives the longitudinal momentum density ($p^+$ density) in the transverse impact parameter space.  We have evaluated  this momentum density in our model.   For unpolarized nucleon the momentum density is axially  symmetric whereas  for a transversely polarized nucleon, the deviation from the axially symmetric distribution is found to be dipolar in nature.

In Sec.\ref{model}, we describe the model very briefly before providing the results for the GFFs  in Sec.\ref{gff}.  In Sec.\ref{PL}, we derive the matrix elements of the energy momentum tensor and the Pauli-Lubanski operator for a transversely polarized proton state with transverse momentum $P^\perp=0$. The sum rule for the intrinsic transverse spin obtained from the  Pauli-Lubanski operator involves the GFFs $A(0),~B(0)$ and $\bar{C}(0)$. In Sec.\ref{sum}, we summarize our main results.  In Sec.\ref{density}, the longitudinal momentum densities in the transverse impact parameter space for both unpolarized and transversely polarized nucleons are discussed. The detail expressions of the matrix elements of $T^{\mu\nu}$ are provided in the appendix.

\section {Light-front quark-diquark model for the nucleon}\label{model}
In quark-scalar diquark model, the nucleon with three valence quarks is considered as an effectively composite system  of a fermion and a neutral scalar bound state of diquark based on one loop quantum fluctuations.  
The generic ansatz for the massless LFWFs as proposed in \cite{Gutsche} is 
\be 
\psi_{+q}^+(x,\bfk) &=&  \varphi_q^{(1)}(x,\bfk) 
\,, \nonumber\\
\psi_{-q}^+(x,\bfk) &=& -\frac{k^1 + ik^2}{xM_n}   \, \varphi_q^{(2)}(x,\bfk) \,, \nonumber\\
\psi_{+q}^-(x,\bfk) &=& \frac{k^1 - ik^2}{xM_n}  \, \varphi_q^{(2)}(x,\bfk)
\,, \nonumber\\
\psi_{-q}^-(x,\bfk) &=& \varphi_q^{(1)}(x,\bfk) 
\,, \label{WF} 
\ee 
where $\psi_{\lambda_q q}^{\lambda_N}(x,\bfk)$ are the LFWFs with specific nucleon helicities $\lambda_N=\pm$ and the struck quark $q$ has a spin   $\lambda_q=\pm$, where plus and minus correspond to $+\frac{1}{2}$ and $-\frac{1}{2}$ respectively.  For the nucleons, $q$ can be either up ($u$) or down($d$) quark.
The functions $\varphi_q^{(1)}(x,\bfk) $ and $\varphi_q^{(2)}(x,\bfk) $ are the wave functions predicted by soft-wall AdS/QCD\cite{BT}
\be
\varphi_q^{(i)}(x,\bfk)=N_q^{(i)}\frac{4\pi}{\kappa}\sqrt{\frac{\log(1/x)}{1-x}}x^{a_q^{(i)}}(1-x)^{b_q^{(i)}}\exp\bigg[-\frac{\bfk^2}{2\kappa^2}\frac{\log(1/x)}{(1-x)^2}\bigg].
\ee
\begin{table}[ht]
\centering 
\begin{tabular}{ c c c } 
\hline\hline 
Parameters&~~~~~~~~~~~~$u$~~~~~~~~~~~~~& ~~~~~~~~~~~$d$~~~~~~~~~~~~  \\ [0.5ex] 
\hline 
$a^{(1)}$ & 0.035 & 0.20\\ 
$b^{(1)}$ & 0.080 & 1.00 \\
\hline
$a^{(2)}$ & 0.75 & 1.25 \\ 
$b^{(2)}$ & -0.60 & -0.20 \\
\hline
$N^{(1)}$ & 29.180 & 33.918 \\
$N^{(2)}$ & 1.459 & 1.413 \\
\hline\hline 
\end{tabular} 
\caption{\label{tab}List of the parameters used in the LF quark-diquark model for $\kappa=406.6~ MeV$ } 
\end{table} 
The normalizations  of  the Dirac and Pauli form factors are fixed as
\be
F_1^q(Q^2)=n_q\frac{I_1^q(Q^2)}{I_1^q(0)},~~~~~~~~~F_2^q(Q^2)=\kappa_q\frac{I_2^q(Q^2)}{I_2^q(0)},\label{D_P_FF}
\ee
so that $F_1^q(0)=n_q$  and $F_2^q(0)=\kappa_q$ where $n_u=2,~n_d=1$ and the anomalous magnetic moments for the $u$ and $d$ quarks are $\kappa_u=1.673$ and $\kappa_d=-2.033$.  The structure integrals, $I_i^q(Q^2)$ obtained from the LFWFs have the form as
\be
I_1^q(Q^2)&=& \int_0^1 dx x^{2a^{(1)}_q} (1-x)^{1+2b^{(1)}_q}R_q(x,Q^2)\exp \bigg[-\frac{Q^2}{4\kappa^2}\log(1/x)\bigg],\label{i1}\\
I_2^q(Q^2)&=& 2\int_0^1 dx x^{2a^{(1)}_q-1} (1-x)^{2+2b^{(1)}_q}\sigma_q(x)\exp \bigg[-\frac{Q^2}{4\kappa^2}\log(1/x)\bigg],\label{i2}
\ee
with
\be
R_q(x,Q^2)&=&1+\sigma_q^2(x)\frac{(1-x)^2}{x^2}\frac{\kappa^2}{M_n^2\log(1/x)}\bigg[1-\frac{Q^2}{4\kappa^2}\log(1/x)\bigg],\label{Rq}\\
\sigma_q(x)&=&\frac{N_q^{(2)}}{N_q^{(1)}}x^{a^{(2)}_q - a^{(1)}_q}(1-x)^{b^{(2)}_q-b^{(1)}_q}.\label{Sq}
\ee
 In this model, the value of the AdS/QCD parameter $\kappa$ is taken to be $406.6 ~MeV$ and the other parameters are fixed by fitting to the electromagnetic properties for the proton and neutron such as form factors, magnetic moments and charge radii \cite{MC}. For completeness, the parameters are listed in the Table \ref{tab}. Here, we should mention that the value of the parameter $\kappa$ depends on exact AdS/QCD model,
here we use the value of $\kappa=406.6~ MeV$ as determined by fitting the nucleon form factors with experimental data  in Ref.\cite{CM}.
\section{Gravitational form factors}\label{gff}
\begin{figure}[htbp]
\begin{minipage}[c]{0.98\textwidth}
\small{(a)}
\includegraphics[width=7.5cm,height=5.5cm,clip]{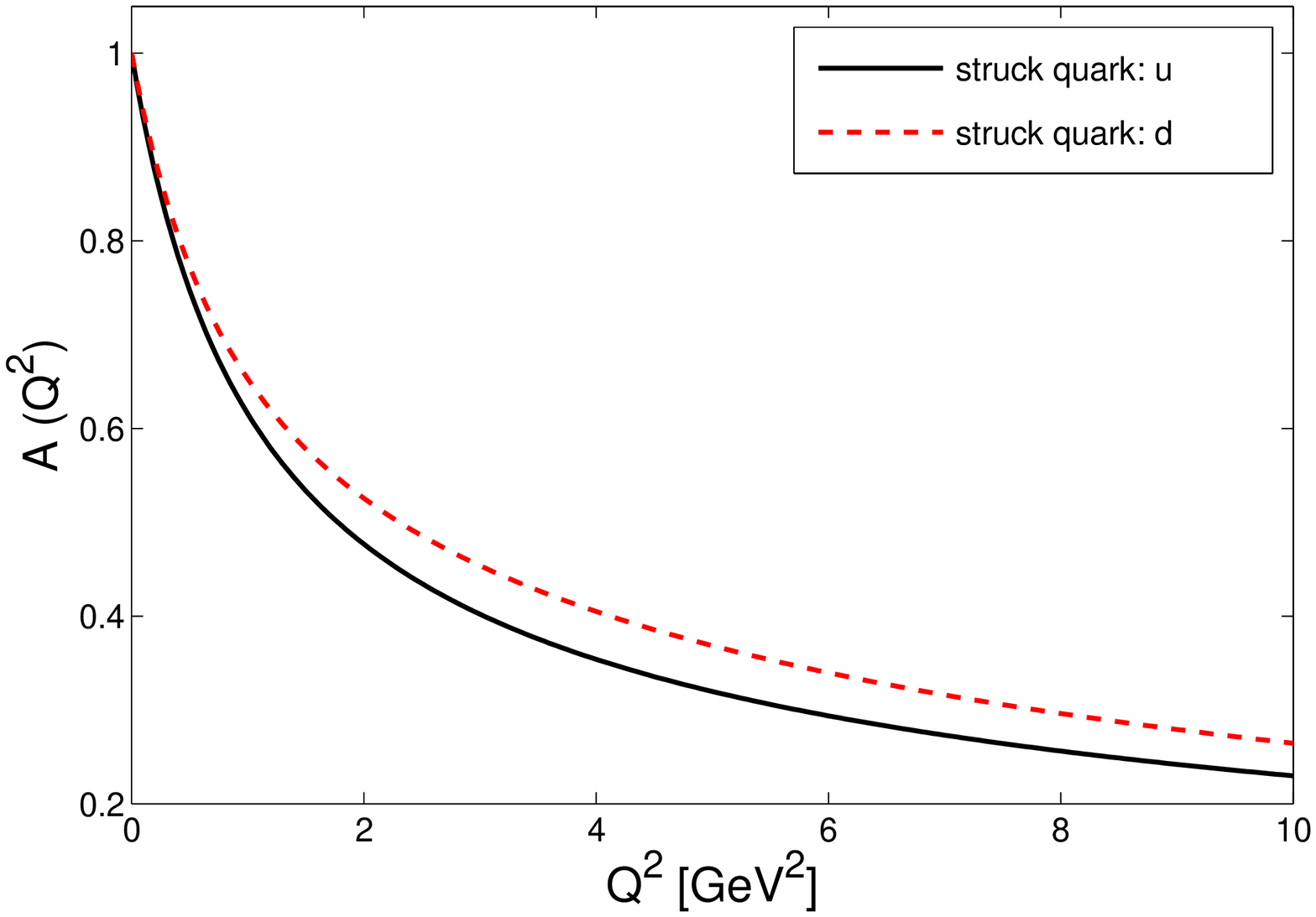}
\hspace{0.1cm}%
\small{(b)}\includegraphics[width=7.5cm,height=5.5cm,clip]{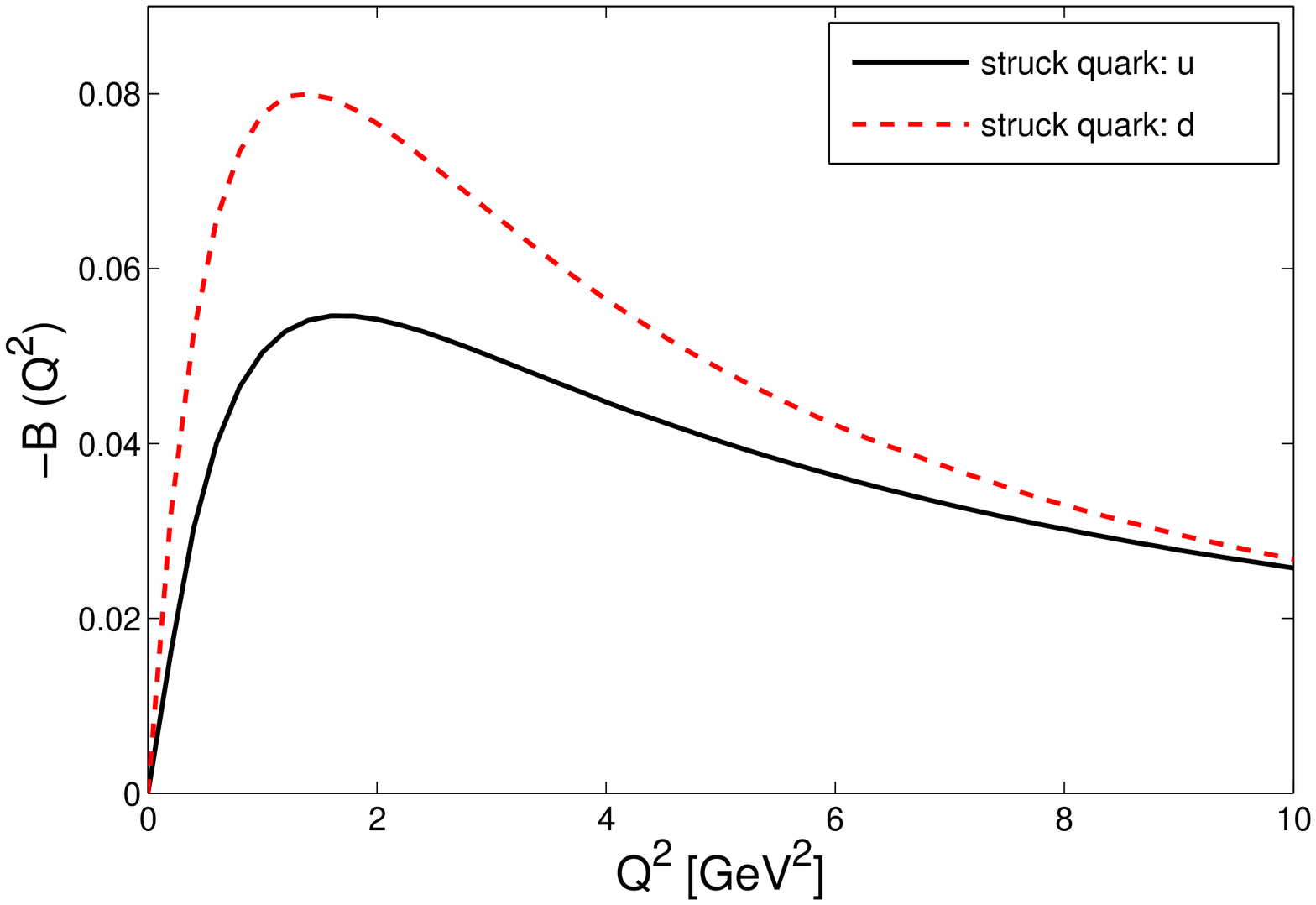}
\end{minipage}
\begin{minipage}[c]{0.98\textwidth}
\small{(c)}\includegraphics[width=7.5cm,height=5.5cm,clip]{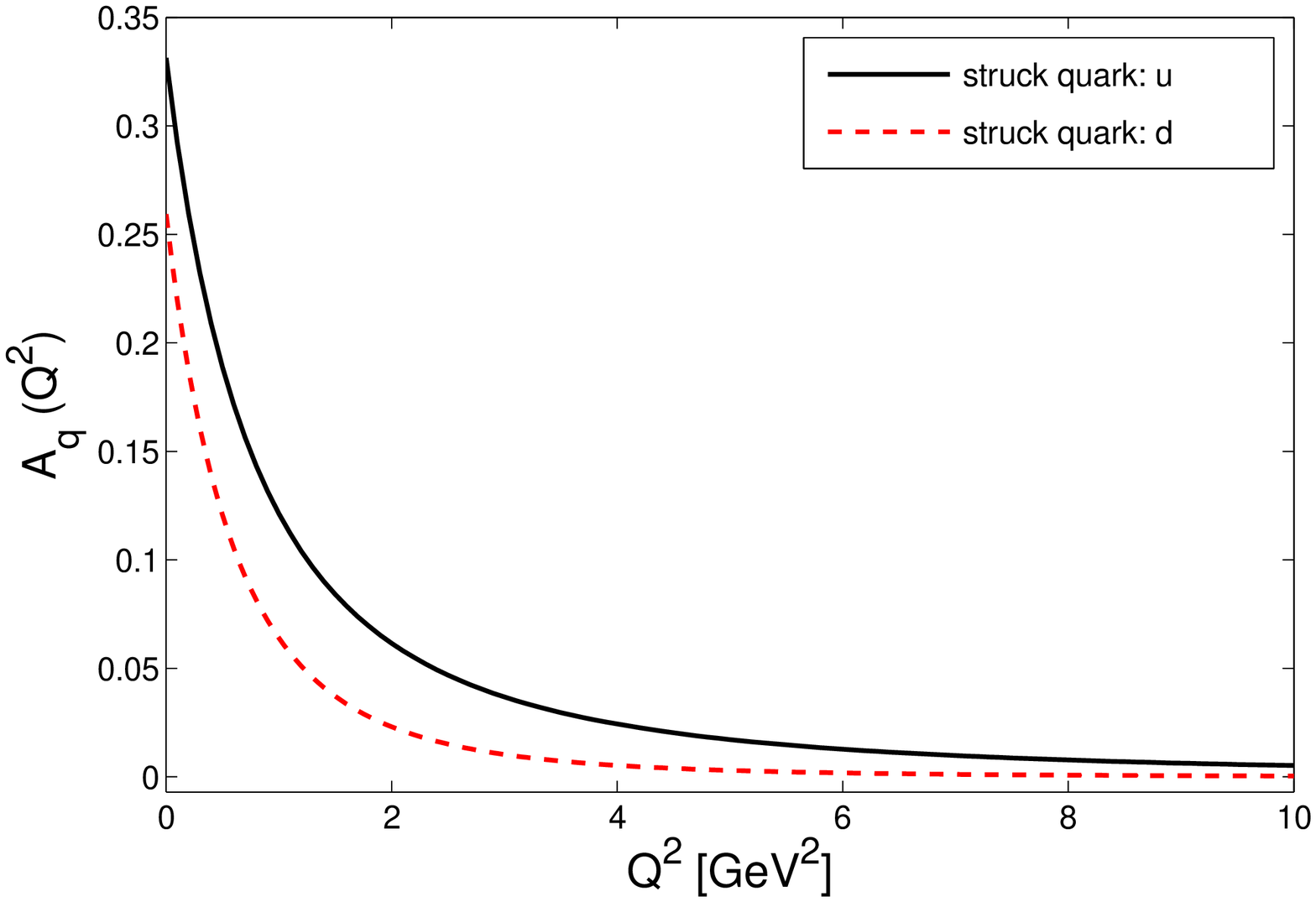}
\hspace{0.1cm}%
\small{(d)}\includegraphics[width=7.5cm,height=5.5cm,clip]{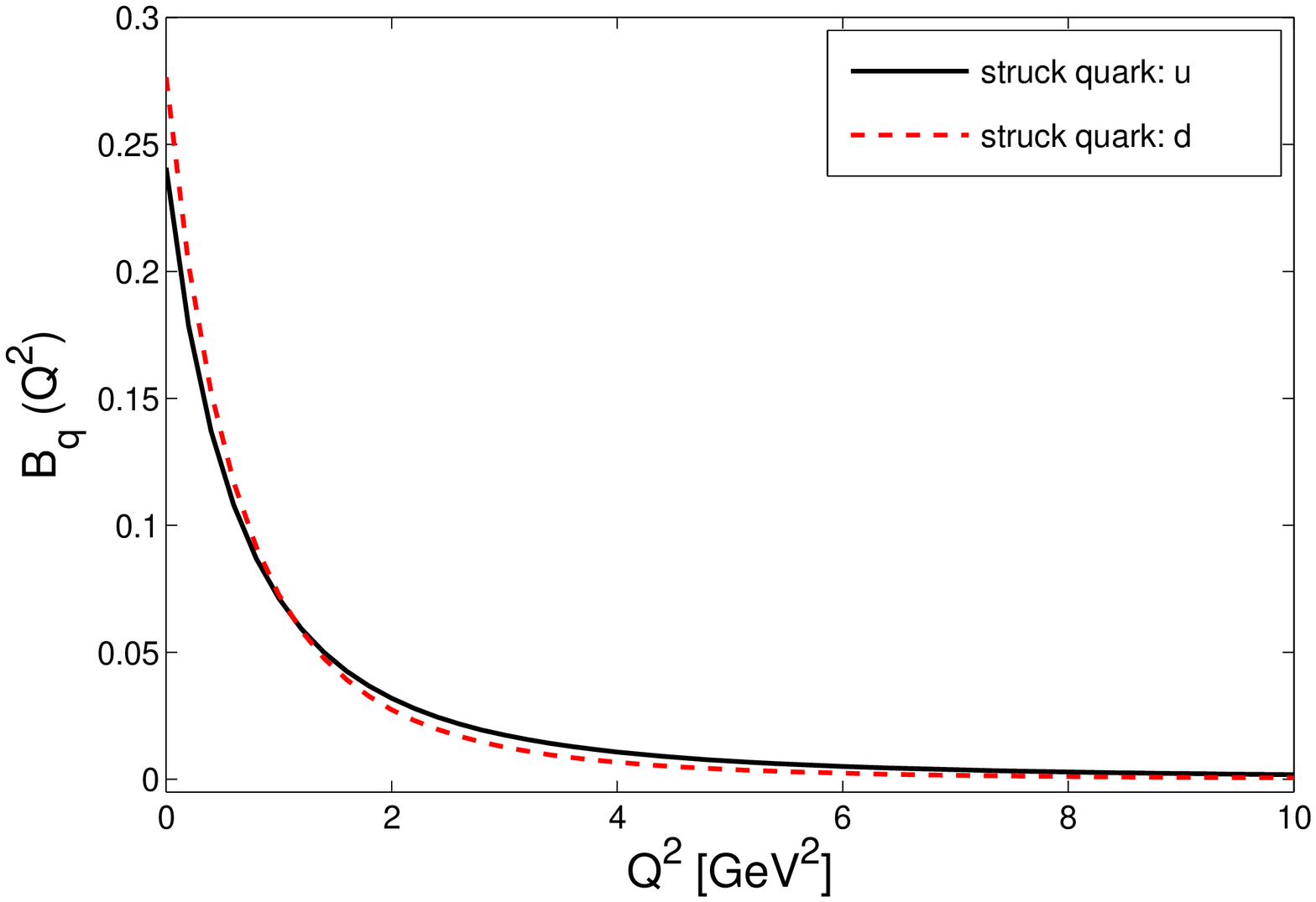}
\end{minipage}
\begin{minipage}[c]{0.98\textwidth}
\small{(e)}\includegraphics[width=7.5cm,height=5.5cm,clip]{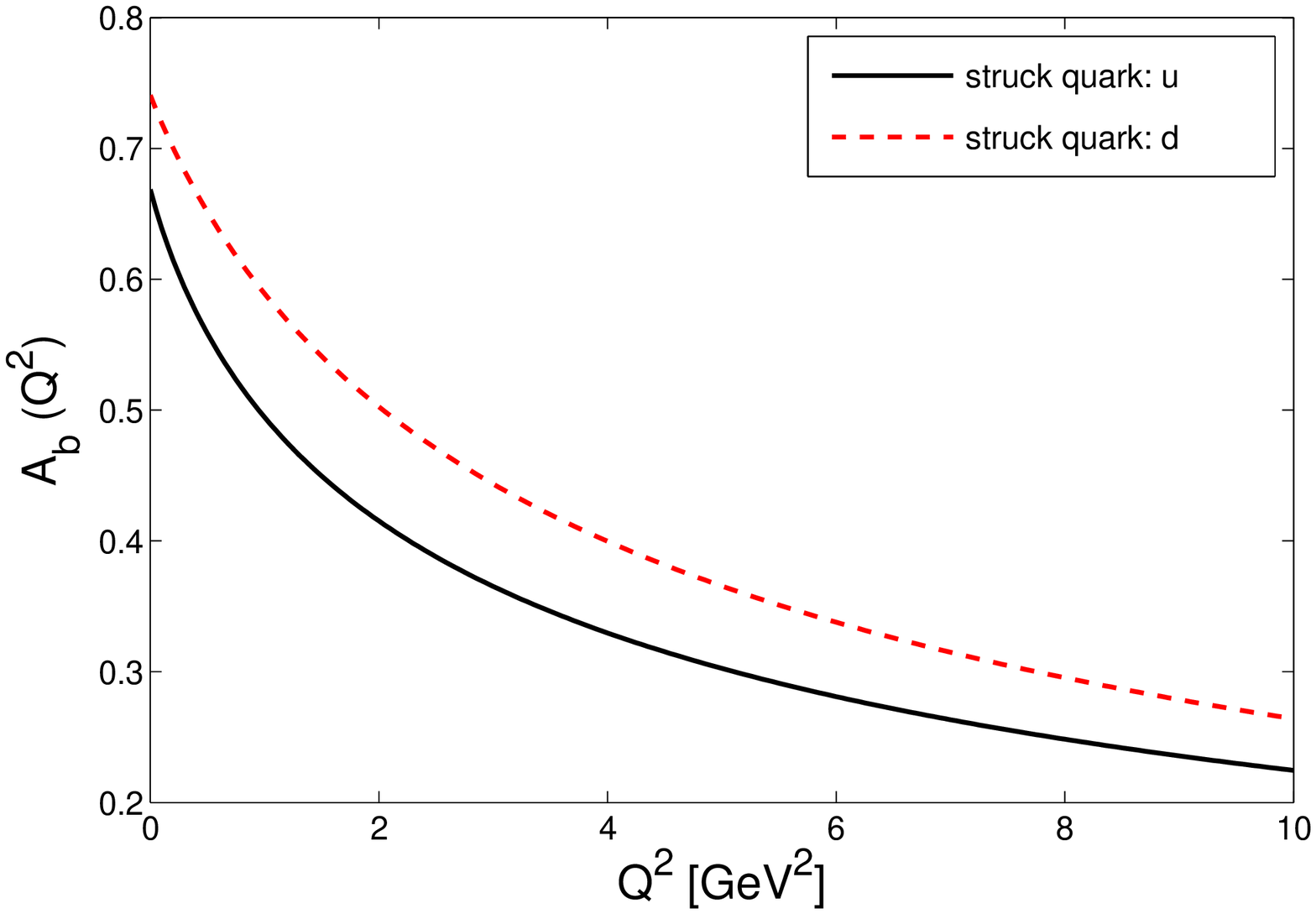}
\hspace{0.1cm}%
\small{(f)}\includegraphics[width=7.5cm,height=5.5cm,clip]{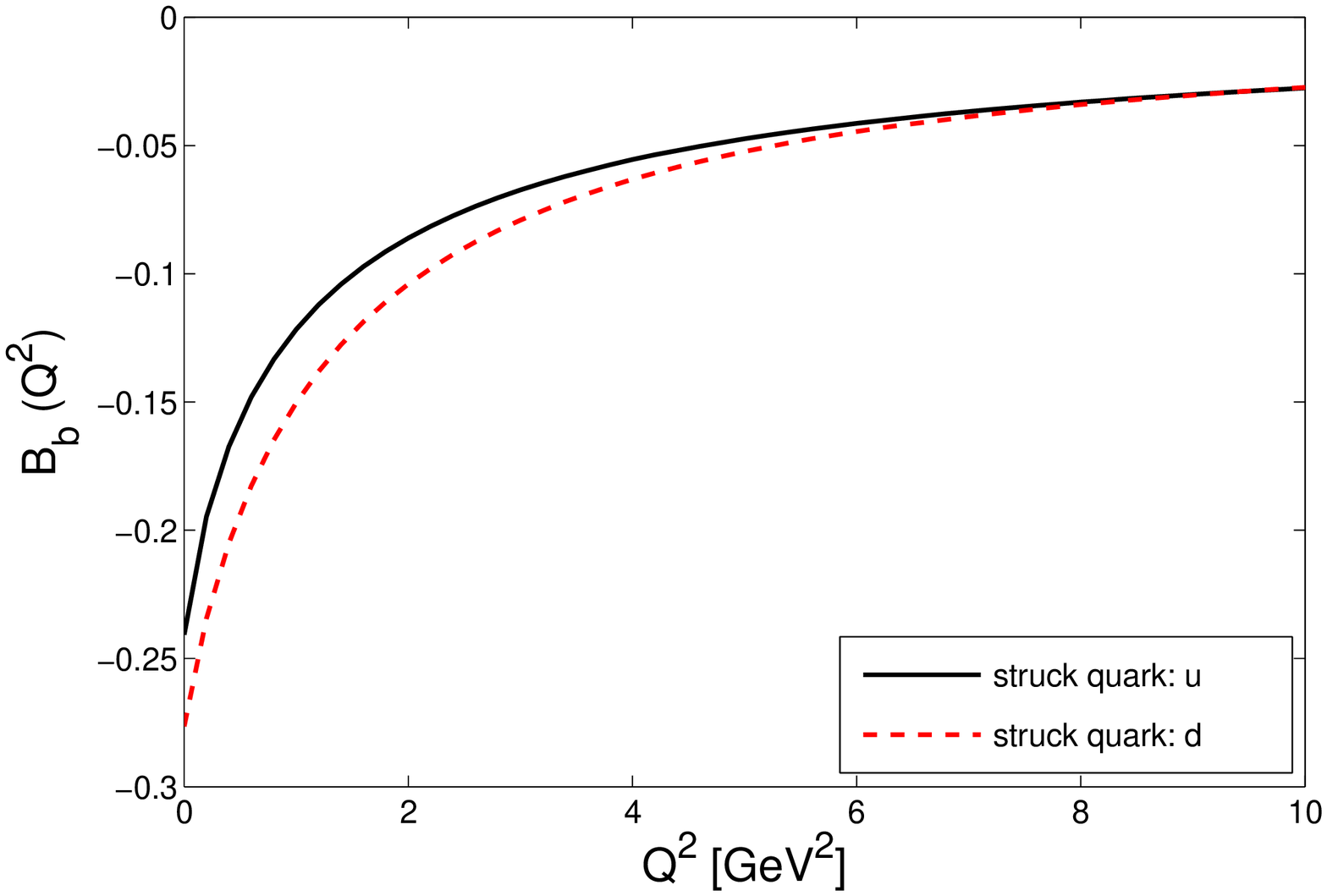}
\end{minipage}
\caption{\label{AB}(Color online) Plots of gravitational form factors $A(Q^2)$ and $B(Q^2)$ for both struck quarks $u$ and $d$. The contributions of quark and diquark to the nucleon are shown in (c)-(f).}
\end{figure}

The gravitational form factors(GFFs) which are related to the matrix elements of the the stress tensor ($T^{\mu \nu}$) play an important role in hadronic physics. For a spin $1/2$ composite system, the matrix elements of $T^{\mu \nu}$ involve four gravitational FFs \cite{hari,ji12} 
\be
\langle P', S'|T^{\mu \nu}_i(0)|P, S\rangle &=&\bar{U}(P', S')\bigg[-B_i(q^2)\frac{\bar{P}^\mu\bar{P^\nu}}{M_n} \nonumber\\ &+&(A_i(q^2)+B_i(q^2))\frac{1}{2}(\gamma^{\mu}\bar{P}^{\nu}+\gamma^{\nu}\bar{P}^{\mu}) \nonumber\\
&+& C_i(q^2)\frac{q^{\mu}q^{\nu}-q^2 g^{\mu\nu}}{M_n}+\bar{C}_i(q^2)M_n g^{\mu\nu}\bigg]U(P,S),\label{tensor}
\ee
where $\bar{P}=(P+P')/2$ and $q=P'-P$ and $A(q^2)$, $B(q^2)$, $C(q^2)$ and $\bar{C}(q^2)$ are the GFFs. The spin-nonflip form factor $A$ is analog of the Dirac form factor $F_1$. $A(q^2)$ allows us to measure the momentum fractions carried by each constituent of a hadron. 
According to Ji's sum rule\cite{ji97}, $2\langle J_q\rangle=A_q(0)+B_q(0)$. Thus, one has to measure the spin-flip
form factor $B$ to find the quark contributions to the nucleon spin. $B(q^2)$ is analogous to the Pauli form factor $F_2$ for the vector current. In the light-front representation, one can easily compute spin-nonflip and spin-flip GFFs by calculating the $++$ component of the matrix elements of the the stress tensor \cite{BH} as
\be
\langle P+q, \uparrow|\frac{T^{++}_i(0)}{2(P^+)^2}|P, \uparrow\rangle &=&A_i(q^2),\\
\langle P+q, \uparrow|\frac{T^{++}_i(0)}{2(P^+)^2}|P, \downarrow\rangle &=&-(q^1-iq^2)\frac{B_i(q^2)}{2M}.
\ee
Here we consider the Yukawa Lagrangian
\begin{equation}
\mathcal{L}=\frac{i}{2}[\bar{\psi}\gamma^{\mu}(\partial_{\mu}\psi)-(\partial_{\mu}\bar{\psi})\gamma^{\mu}\psi]-m\bar{\psi}\psi+\frac{1}{2}(\partial^{\mu}\phi)(\partial_{\mu}\phi)-\frac{1}{2}\lambda^2\phi\phi+g\phi\bar{\psi}\psi \,  ,
\end{equation}
which leads to the corresponding energy momentum tensor as 
\begin{equation}
T^{\mu\nu}=\frac{i}{2}[\bar{\psi}\gamma^{\mu}(\overrightarrow{\partial}^{\nu}\psi)-\bar{\psi}\gamma^{\mu}\overleftarrow{\partial}^{\nu}\psi]+(\partial^{\mu}\phi)(\partial^{\nu}\phi)-g^{\mu\nu}\mathcal{L} \,  .
\end{equation}
Using the two particle Fock states for $J^z=+\frac{1}{2}$ and $J^z=-\frac{1}{2}$ and the light-front wave functions given in Eq.(\ref{WF}), we evaluate the GFFs $A(q^2)$ and $B(q^2)$ depending on different flavors (struck quark) as
\eq
 A(q^2)&=& A^q(q^2)+A^b(q^2)=\frac{\mathcal{I}_{1q}(q^2)+\mathcal{I}_{1b}(q^2)}{I_1^q(0)} \,,\label{A}\\
 B(q^2)&=& B^q(q^2)+B^b(q^2)=2M_n\frac{\mathcal{I}_{2q}(q^2)-\mathcal{I}_{2b}(q^2)}{I_2^q(0)} \label{B}\,,
\en
where $A^{q/b}(q^2)$ and $\mathcal{I}_i^{q/b}(q^2)$ are the GFFs and structure integrals corresponding to quark/scalar diquark.  The explicit expressions of the structure integrals are listed in the Appendix.
The integrals, $I_i^q(Q^2)$ in the denominators in the right hand side of  Eq.(\ref{A}) and Eq.(\ref{B}) have the form as
\be
I_1^q(Q^2)&=& \int_0^1 dx x^{2a^{(1)}_q} (1-x)^{1+2b^{(1)}_q}R_q(x,Q^2)\exp \bigg[-\frac{Q^2}{4\kappa^2}\log(1/x)\bigg],\\
I_2^q(Q^2)&=& 2\int_0^1 dx x^{2a^{(1)}_q-1} (1-x)^{2+2b^{(1)}_q}\sigma_q(x)\exp \bigg[-\frac{Q^2}{4\kappa^2}\log(1/x)\bigg],
\ee
with
$ R_q(x,Q^2) $ and $\sigma_q(x) $  as defined in Eqs.(\ref{Rq},\ref{Sq}).
 In Fig. \ref{AB}(a) and (b), we show total $A(q^2)$ and $B(q^2)$ depending on different struck quarks. The contributions of quark and the diquark to the total spin-nonflip and spin-flip GFFs are shown in Fig. \ref{AB}(c)-(f). One notices that at zero momentum transfer, $A(0)=A^q(0)+A^b(0)=1$ and $B(0)=B^q(0)+B^b(0)=0$ as expected. 
\section{Matrix elements of the energy-momentum tensor}\label{PL}
 Here we consider the transversely polarized state to calculate the matrix elements of $T^{\mu\nu}$. The transversely polarized state (polarized along +ve $x$ direction)  is given by
\eq
|P,S^{(1)}\rangle=\frac{1}{\sqrt{2}}\bigg(|\Psi_{2p}^{\uparrow}\big(P^+,P^{\perp}\big)\rangle+|\Psi_{2p}^{\downarrow}\big(P^+,P^{\perp}\big)\rangle \bigg) \, ,
\en
where $\Psi_{2p}^{\uparrow}(\Psi_{2p}^{\downarrow})$ represents the two-particle Fock state corresponding to $J^z=+\frac{1}{2}(J_z=-\frac{1}{2})$.   For the transversely polarized state we calculate the matrix elements of $T^{\mu\nu}$ for $\Psi_{2p}^{\uparrow}$ going to $\Psi_{2p}^{\downarrow}$ and $\Psi_{2p}^{\downarrow}$ going to $\Psi_{2p}^{\uparrow}$.
 To evaluate the right hand side of Eq.(\ref{tensor}), we use the matrix elements of the different $\gamma$ matrices listed in the appendix of the Ref. \cite{hari}. Here we list only the final expressions of all the matrix elements for zero skewness and the detailed calculations are given in the appendix. 
\eq
&&\langle \Psi_{2p}^{\uparrow}\big(P'\big)|T^{++}|\Psi_{2p}^{\downarrow}\big(P\big)\rangle + \langle \Psi_{2p}^{\downarrow}\big(P'\big)|T^{++}|\Psi_{2p}^{\uparrow}\big(P\big)\rangle \nonumber\\
&=&2(P^+)^2\frac{2(\mathcal{I}_{2q}-\mathcal{I}_{2b})}{I_2^q(0)}(iq^2_{\perp})
=B(Q^2)\frac{2(P^+)^2}{M_n}(iq^2_{\perp}),
\en
\eq
&&{\langle \Psi_{2p}^{\uparrow}\big(P'\big)|T^{+1}|\Psi_{2p}^{\downarrow}\big(P\big)\rangle + \langle \Psi_{2p}^{\downarrow}\big(P'\big)|T^{+1}|\Psi_{2p}^{\uparrow}\big(P\big)\rangle }\nonumber\\
&=&\frac{2(\mathcal{I}_{4q}-\mathcal{I}_{4b})}{I_2^q(0)}P^+ (iq^1_{\perp}q^2_{\perp})=B(Q^2)\frac{P^+}{M_n}(iq^1_{\perp}q^2_{\perp}) \,.
\en
\eq
&&{\langle \Psi_{2p}^{\uparrow}\big(P'\big)|T^{+2}|\Psi_{2p}^{\downarrow}\big(P\big)\rangle + \langle \Psi_{2p}^{\downarrow}\big(P'\big)|T^{+2}|\Psi_{2p}^{\uparrow}\big(P\big)\rangle }\nonumber\\
&=&\frac{2(\mathcal{I}_{6q}-\mathcal{I}_{6b})}{I_2^q(0)}P^+ i(q^2_{\perp})^2=B(Q^2)\frac{P^+}{M_n}i(q^2_{\perp})^2 \,.
\en
\eq
&&{\langle \Psi_{2p}^{\uparrow}\big(P'\big)|T^{+-}|\Psi_{2p}^{\downarrow}\big(P\big)\rangle + \langle \Psi_{2p}^{\downarrow}\big(P'\big)|T^{+-}|\Psi_{2p}^{\uparrow}\big(P\big)\rangle }
=\frac{2(\mathcal{I}_{8q}-\mathcal{I}_{8b})}{I_2^q(0)} (iq^2_{\perp})\nonumber\\
&=&-\big[A(Q^2)(2M_n)-B(Q^2)\frac{(q^{\perp})^2}{M_n}+C(Q^2)\frac{4(q^{\perp})^2}{M_n}+\bar{C}(Q^2)(4M_n)\big](iq^2_{\perp}) \,.
\en
\begin{figure}[htbp]
\begin{minipage}[c]{0.98\textwidth}
\small{(a)}
\includegraphics[width=7.5cm,height=5.5cm,clip]{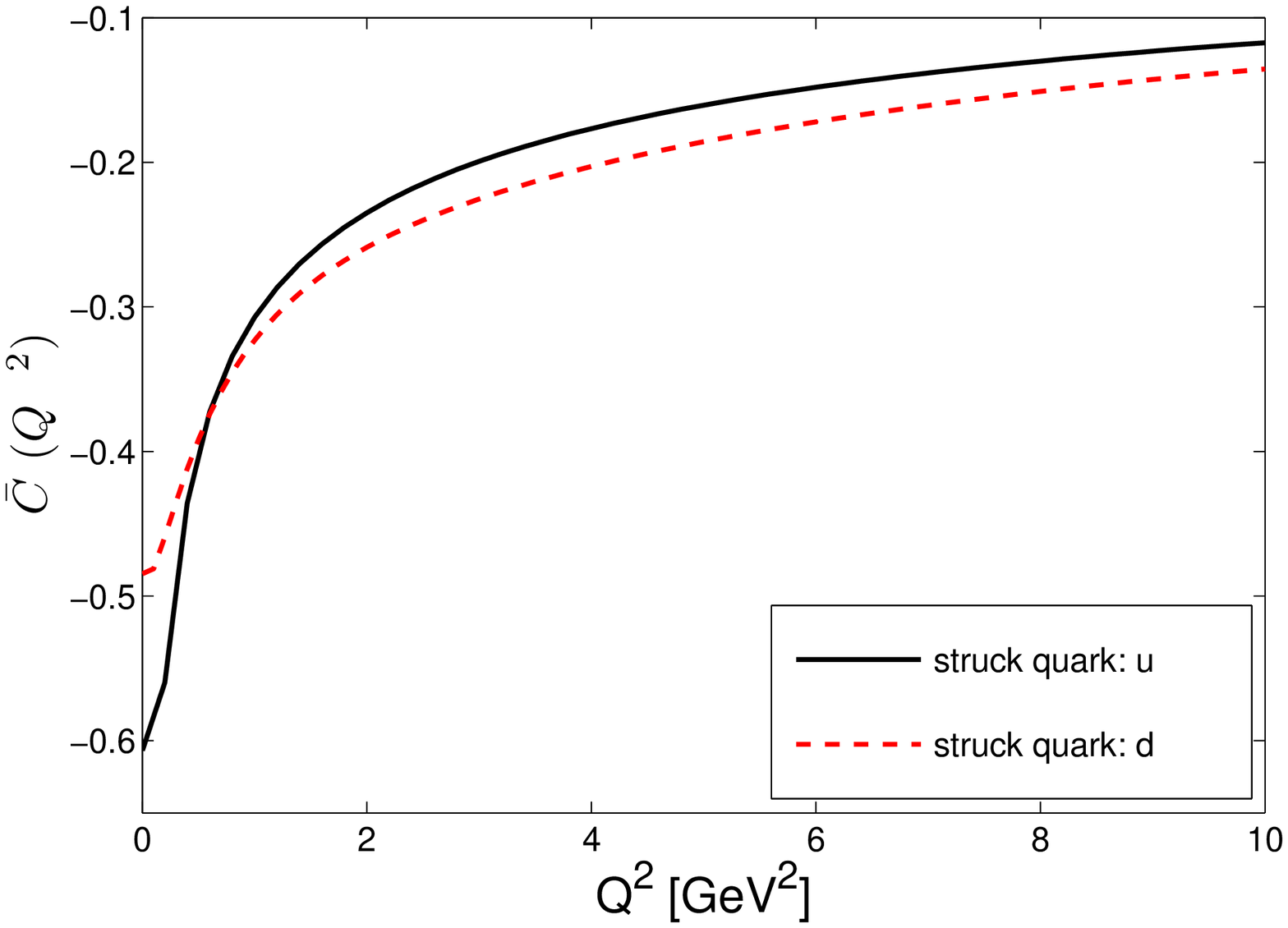}
\end{minipage}
\begin{minipage}[c]{0.98\textwidth}
\small{(b)}\includegraphics[width=7.5cm,height=5.5cm,clip]{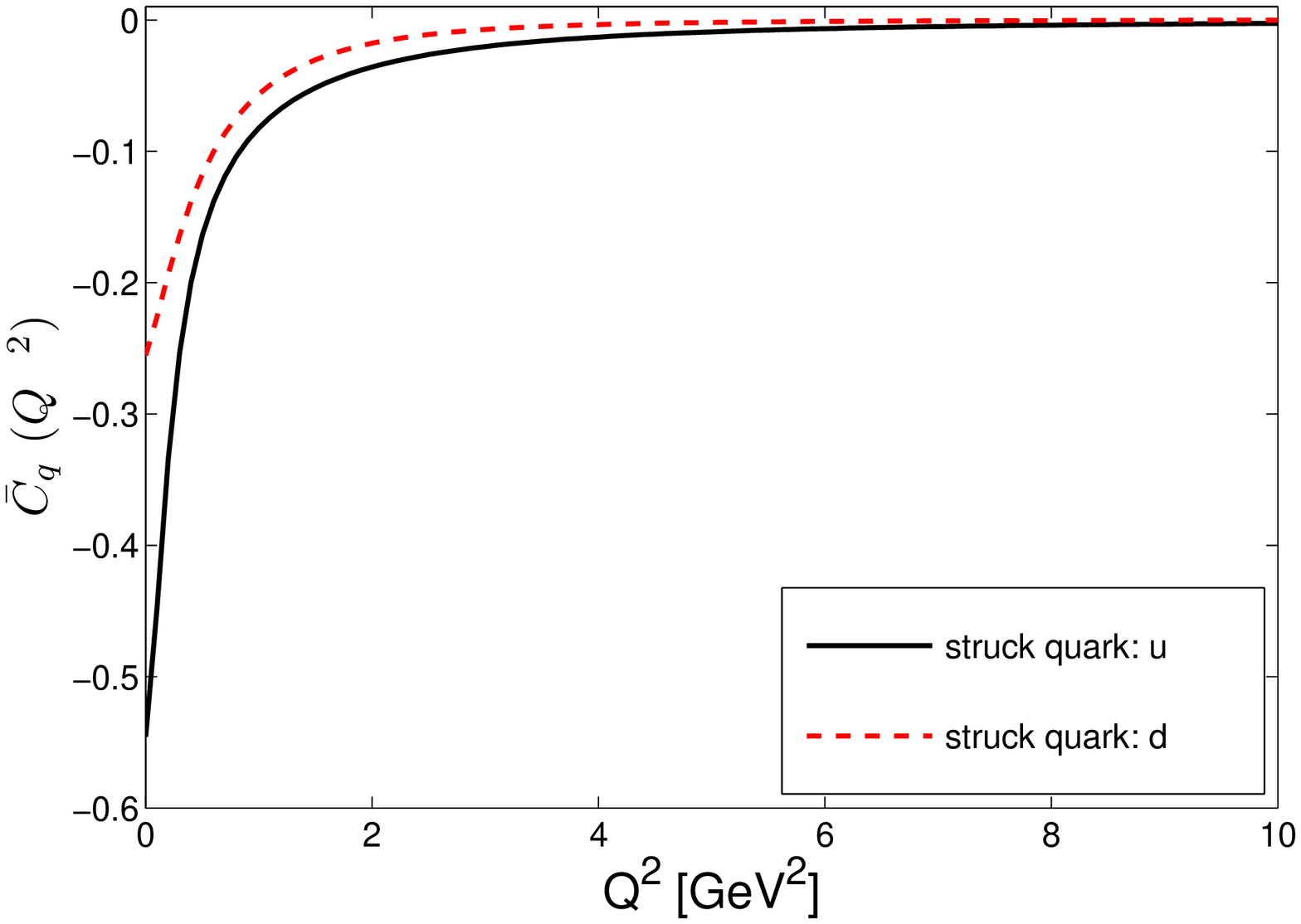}
\small{(c)}\includegraphics[width=7.5cm,height=5.5cm,clip]{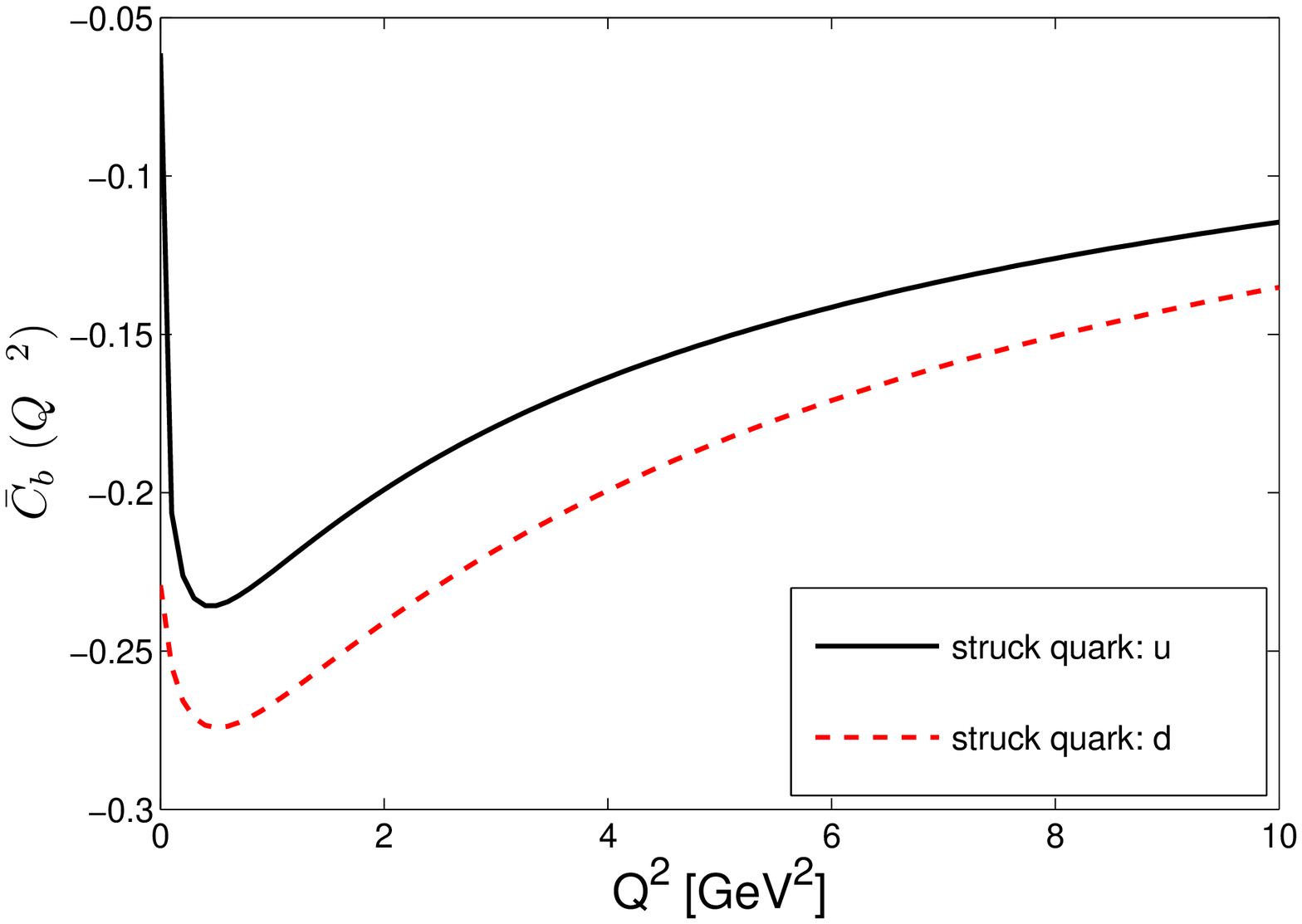}
\end{minipage}
\caption{\label{cbar}(Color online) Plots of gravitational form factor $\bar{C}(Q^2)$ for both struck quarks $u$ and $d$. The contributions of quark and diquark to the nucleon are shown in (b) and (c).}
\end{figure}
 We keep only terms linear in $\bf q$, which are relevant for evaluation of the  matrix elements of transverse spin.  The matrix elements of $T^{\mu\nu}$  up to $\mathcal{O}{(q)}$ are:
\eq
\langle P',S^{(1)}|T^{++}|P,S^{(1)}\rangle &=& (P^+)^2\frac{2(\mathcal{I}_{2q}-\mathcal{I}_{2b})}{I_2^q(0)}(iq^2_{\perp})
=B(Q^2)\frac{(P^+)^2}{M_n}(iq^2_{\perp}),\label{1}\\
\langle P',S^{(1)}|T^{+1}|P,S^{(1)}\rangle &=& 0,\label{2}\\
\langle P',S^{(1)}|T^{+2}|P,S^{(1)}\rangle &=& 0,\label{3}\\
\langle P',S^{(1)}|T^{+-}|P,S^{(1)}\rangle &=& \frac{(\mathcal{I}_{8q}-\mathcal{I}_{8b})}{I_2^q(0)} (iq^2_{\perp})= -M_n\big[A(Q^2)+2\bar{C}(Q^2)\big](iq^2_{\perp}) \label{4}\,.
\en

We see from the Eq.(\ref{1}) that the matrix element of $T^{++}$ in a transversely polarized state does not depend on the form factor $A(Q^2)$. It depends only on the form factor $B(Q^2)$. Whereas, Eq.(\ref{4}) implies that the matrix element of $T^{+-}$ depends on both $\bar{C}(Q^2)$ and  $A(Q^2)$. 
For nonzero skewness, the matrix element of $T^{+2}$ has a term proportional to $q^+$ as shown in \cite{hari}.
 But the matrix element of $T^{+2}$ in this quark-diquark model is zero when we consider only the term linear in $\bf q$. The main reason is that the LFWFs are independent of quark mass in this model. For zero skewness, the results of this quark-diquark model are consistent with Ref. \cite{hari}. Using Eq.(\ref{A}) and (\ref{4}), we evaluate the $\bar{C}(Q^2)$ form factor. In Fig. \ref{cbar} we show the form factor $\bar{C}(Q^2)$ for different struck quark. The quark and the diquark contributions are 
shown in Fig. \ref{cbar}(b) and (c) respectively.   
\subsection{Matrix element of the Pauli-Lubanski operator}
The Pauli-Lubanski operator is defined as \cite{hari}
\be
W^1_i=\frac{1}{2}F_i^2P^++\tilde{K}^3P^2-\frac{1}{2}\tilde{E}_i^2P^-,
\ee
where $F^i$ and $E^i$ are the light front transverse rotation and transverse boost operators. $K^3$ is the longitudinal boost operator. The matrix elements of the operators $F_i^2$, $\tilde{K}^3$ and $\tilde{E}_i^2$ in a transversely polarized state are given by
\be
\langle PS^{(1)}|F_i^2|PS^{(1)}\rangle &=& i(2\pi)^3\delta^3(0)\Big[\frac{\partial}{\partial \Delta_-}\langle P'S^{(1)}|T^{+2}_i(0)|PS^{(1)}\rangle \nonumber\\
&&- \frac{\partial}{\partial \Delta_2^{\perp}}\langle P'S^{(1)}|T^{+-}_i(0)|PS^{(1)}\rangle\Big]_{q=0},\label{mat1}
\ee
\be
\langle PS^{(1)}|\tilde{K}_i^3|PS^{(1)}\rangle=-\frac{i}{2}(2\pi)^3\delta^3(0)\Big[\frac{\partial}{\partial \Delta_-}\langle P'S^{(1)}|T^{++}_i(0)|PS^{(1)}\rangle \Big]_{q=0},\label{mat2}
\ee
and
\be
\langle PS^{(1)}|\tilde{E}_i^2|PS^{(1)}\rangle=-i(2\pi)^3\delta^3(0)\Big[\frac{\partial}{\partial \Delta_2^{\perp}}\langle P'S^{(1)}|T^{++}_i(0)|PS^{(1)}\rangle \Big]_{q=0},\label{mat3}
\ee
where $\Delta=P'-P$. Using the results of the individual matrix element in Eqs. (\ref{mat1}), (\ref{mat2}) and (\ref{mat3}), the matrix element of the total Pauli-Lubanski operator $W^1$ can be written in this quark-diquark model as
\be
\frac{\langle PS^{(1)}|W^1|PS^{(1)}\rangle}{\langle PS^{(1)}|PS^{(1)}\rangle}&=&\frac{\langle PS^{(1)}|W^1|PS^{(1)}\rangle}{(2\pi)^3 2P^+\delta^3(0)}\nonumber\\
&=&\frac{1}{2P^+}\Big[\frac{P^+}{2}\Big\{\frac{\partial}{\partial \zeta}\frac{(\mathcal{I}_{9q}^{I}+\mathcal{I}_{9b}^{I})}{I_2^q(0)}\Big|_{q=0}\Big\}\nonumber\\
&+&\frac{P^+}{2}\Big \{-\frac{\mathcal{I}_{8q}(0)-\mathcal{I}_{8b}(0)}{I_2^q(0)}\Big\}+\frac{P^-}{2}(P^+)^2\bigg\{\frac{2\mathcal{I}_2^q(0)}{I_2^q(0)}-\frac{2\mathcal{I}_2^b(0)}{I_2^q(0)}\bigg\}\Big]\nonumber\\
&=&\frac{1}{2P^+}\Big[\frac{P^+}{2} M_n\{2A(0)+B(0)+2\bar{C}(0)\}+\frac{P^-}{2}(P^+)^2\frac{B(0)}{M_n}\Big]\nonumber\\
&=&\frac{M_n}{2}\Big[A(0)+B(0)+\bar{C}(0)\Big].
\ee
One can notice that as $B(0)=0$ (Fig. \ref{AB}-b), only the matrix element of $T^{+-}$ make contribution to the matrix element of total $W^1$ operator in a transversely polarized state. For different struck quarks, the matrix element of total $W^1$ operator will give different values due as $\bar{C}(0)$ is different for $u$ and $d$ quarks. The non-vanishing contribution of $\bar{C}$ to the matrix element of Pauli-Lubanski operator $W^1$ has been reported previously in \cite{hari,hatta, leader}. 

For a massive particle like nucleon, the intrinsic spin operators can be related with the Pauli-Lubanski operators through the following relations \cite{hari, hari2,hari3},
\be
M_n {\cal J}^i  &=& W^i - P^i {\cal J}^3 
= \epsilon^{ij}(\frac{1}{2} F^j P^+ + K^3 P^j- \frac{1}{2} E^j P^-)- P^i {\cal J}^3 ~,
\nonumber\\
{\cal J}^3 &=& \frac {W^+}{P^+} = J^3 + \frac{1}{P^+}(E^1 P^2 -  E^2 P^1),
\ee
where $J^3$ is the helicity operator. The matrix element of the intrinsic spin operator in a transversely polarized dressed quark state has been explicitly demonstrated in \cite{hari3}. The matrix element of the intrinsic spin operator in a transversely polarized state in the quark-diquark model is given by
\be
\frac{\langle PS^{(1)}|{\cal J}^1|PS^{(1)}\rangle}{\langle PS^{(1)}|PS^{(1)}\rangle}=\frac{1}{M_n}\frac{\langle PS^{(1)}|W^1|PS^{(1)}\rangle}{\langle PS^{(1)}|PS^{(1)}\rangle}=\frac{1}{2}\Big[A(0)+B(0)+\bar{C}(0)\Big].
\ee
In \cite{hari}, the authors have described in detail whether the matrix elements of the intrinsic spin operator are frame independent or not whereas the authors in \cite{ji12,ji3} have claimed that the results are frame independent though they have calculated only in a frame where $P^{\perp}=0$.
\begin{figure}[htbp]
\begin{minipage}[c]{0.98\textwidth}
\small{(a)}
\includegraphics[width=7cm,height=5.5cm,clip]{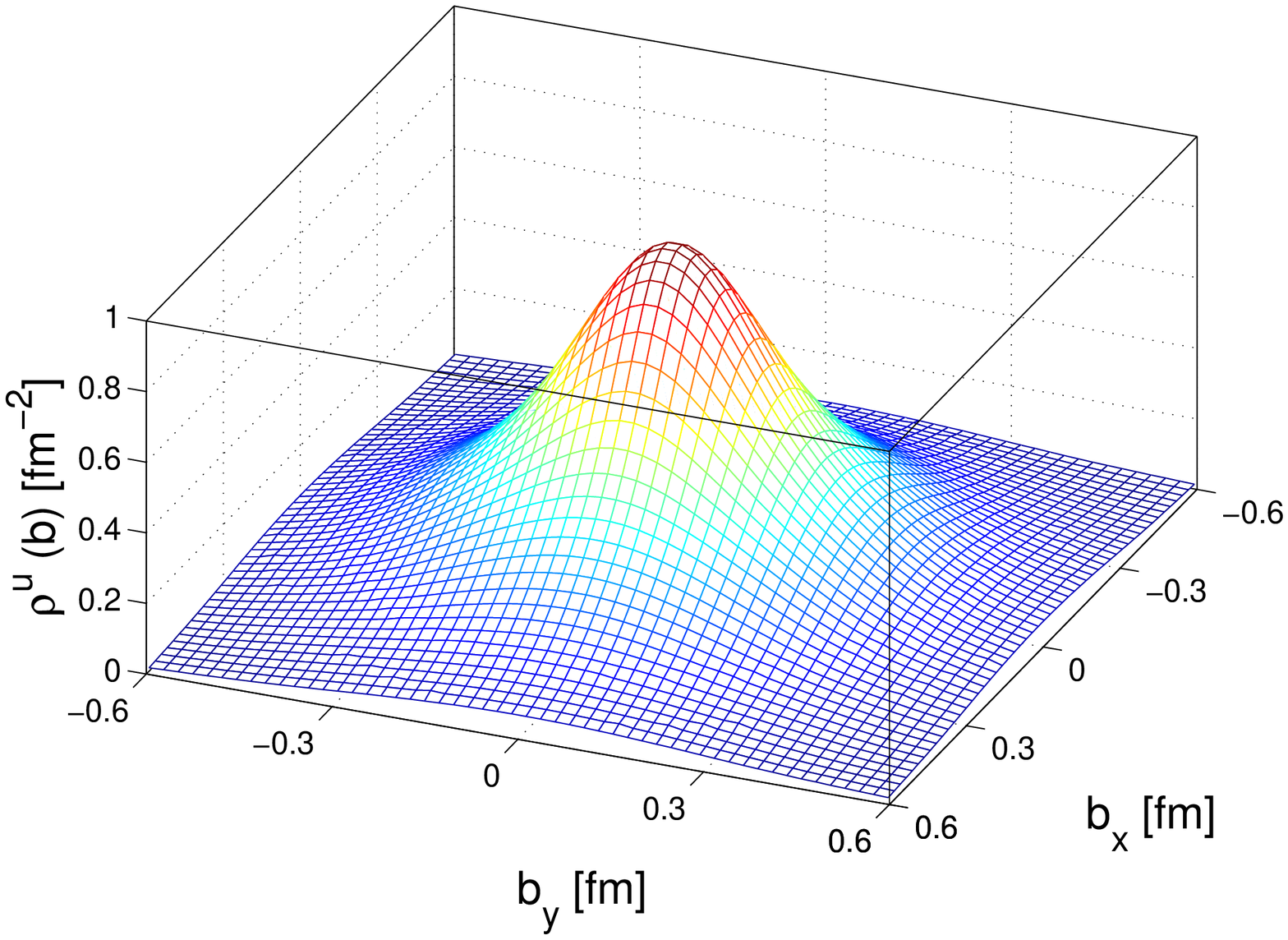}
\hspace{0.1cm}%
\small{(b)}\includegraphics[width=7cm,height=5.5cm,clip]{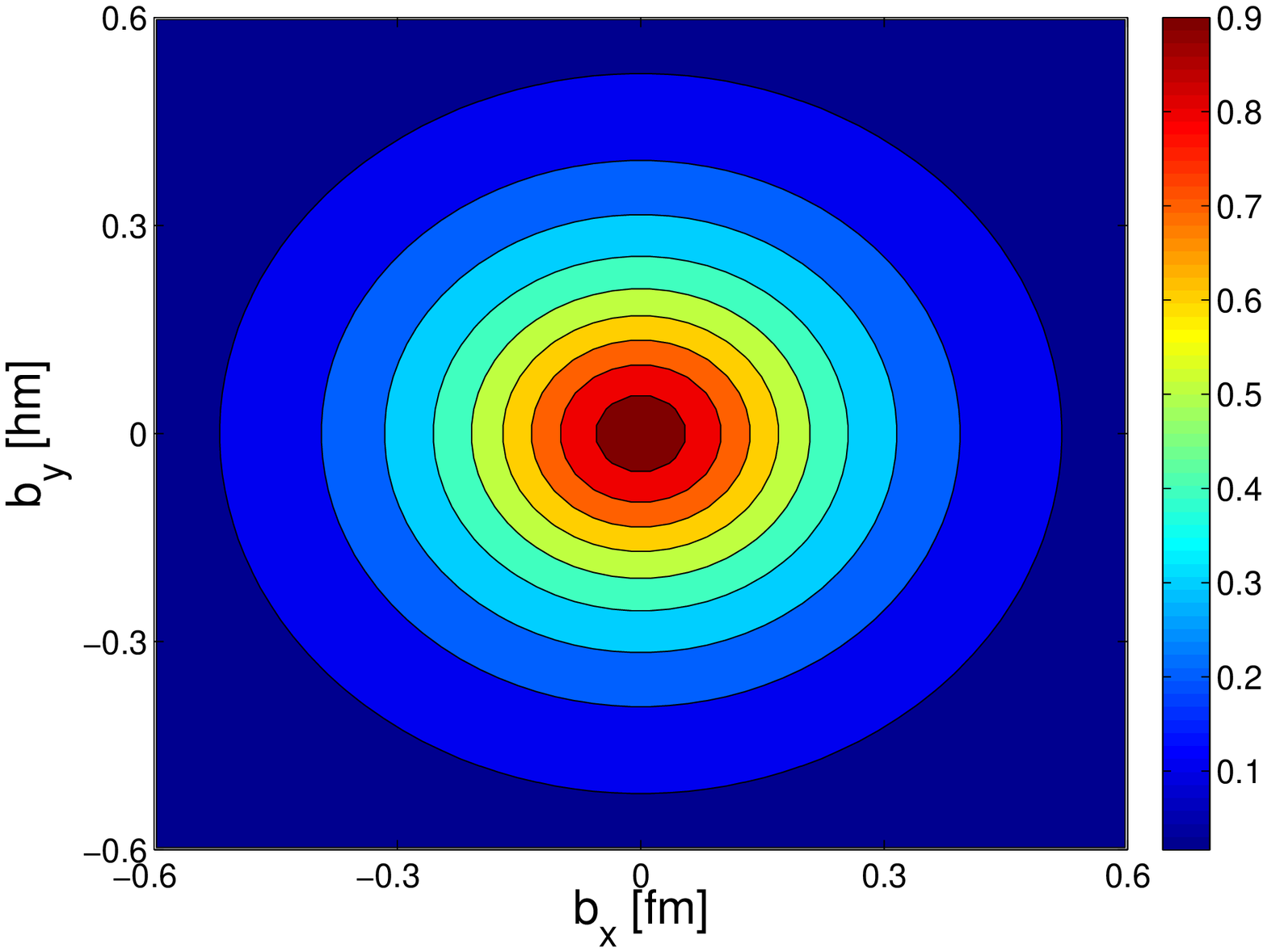}
\end{minipage}
\begin{minipage}[c]{0.98\textwidth}
\small{(c)}
\includegraphics[width=7cm,height=5.5cm,clip]{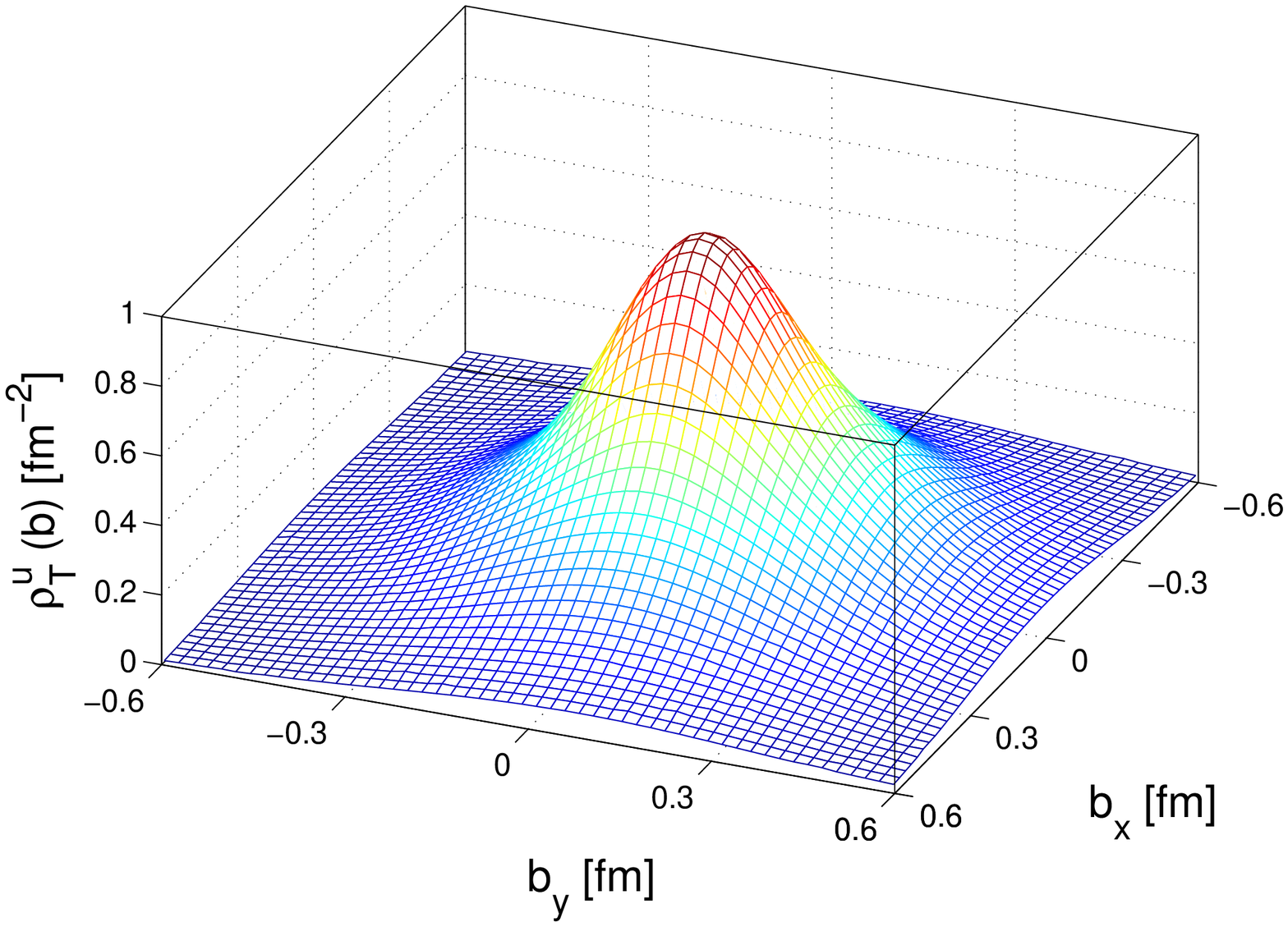}
\hspace{0.1cm}%
\small{(d)}\includegraphics[width=7cm,height=5.5cm,clip]{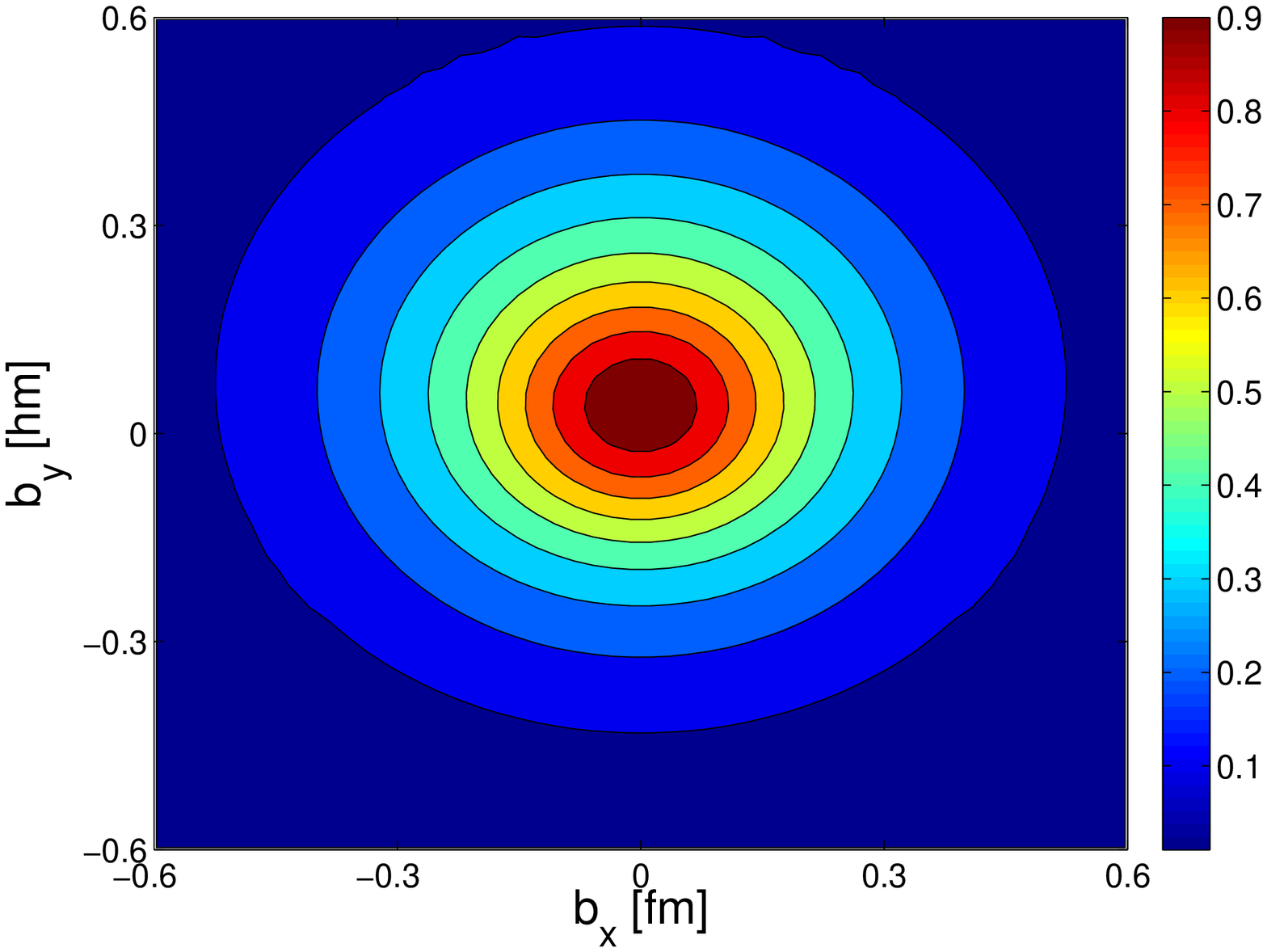}
\end{minipage}
\caption{\label{den_u}(Color online) The longitudinal momentum densities for the active $u$ quark in the transverse plane, upper panel  for unpolarized nucleon, lower panel for  nucleon  polarized along $x$-direction. (b) and (d) are the top view of (a) and (c) respectively.}
\end{figure} 
\begin{figure}[htbp]
\begin{minipage}[c]{0.98\textwidth}
\small{(a)}
\includegraphics[width=7cm,height=5.5cm,clip]{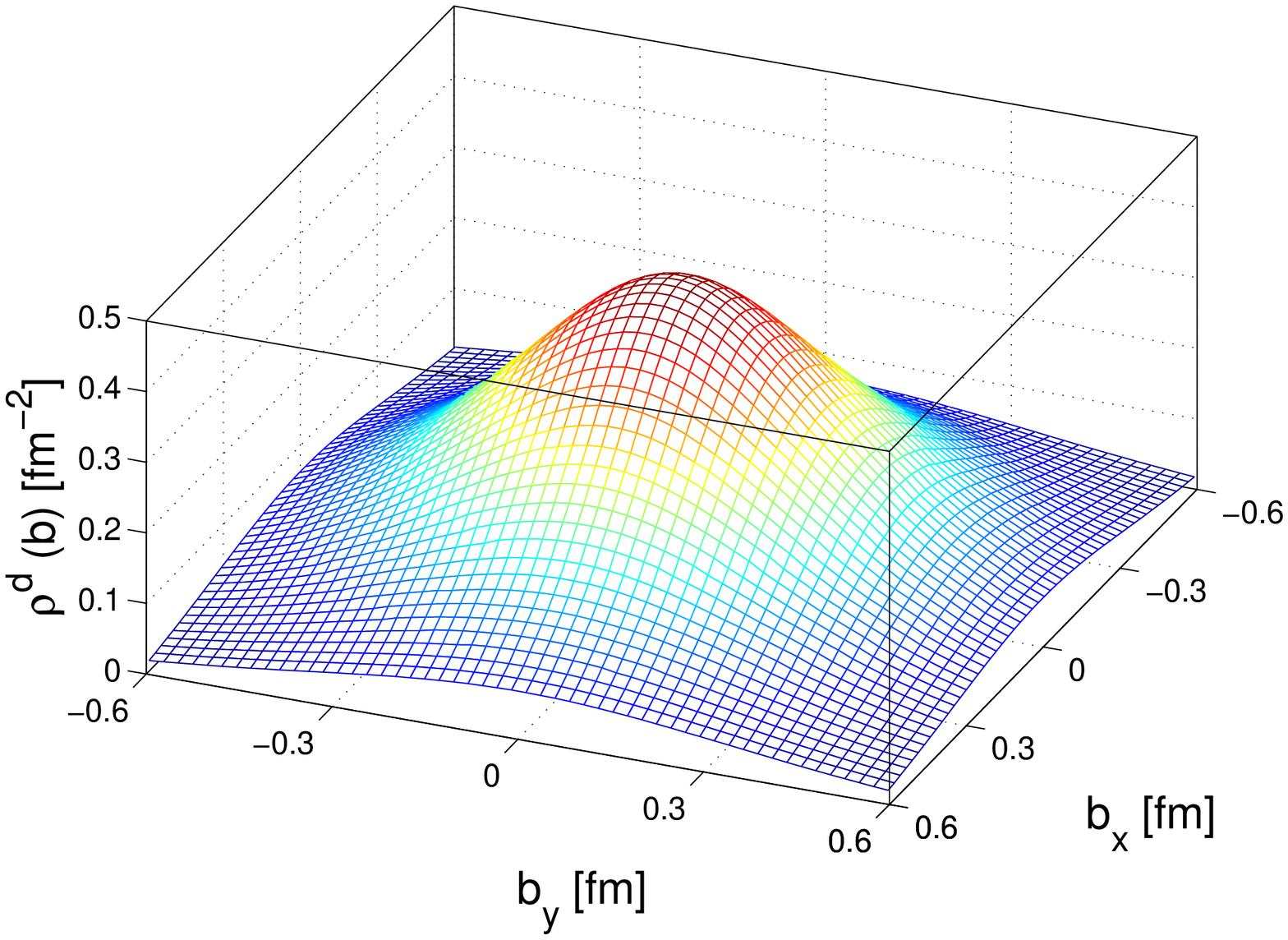}
\hspace{0.1cm}%
\small{(b)}\includegraphics[width=7cm,height=5.5cm,clip]{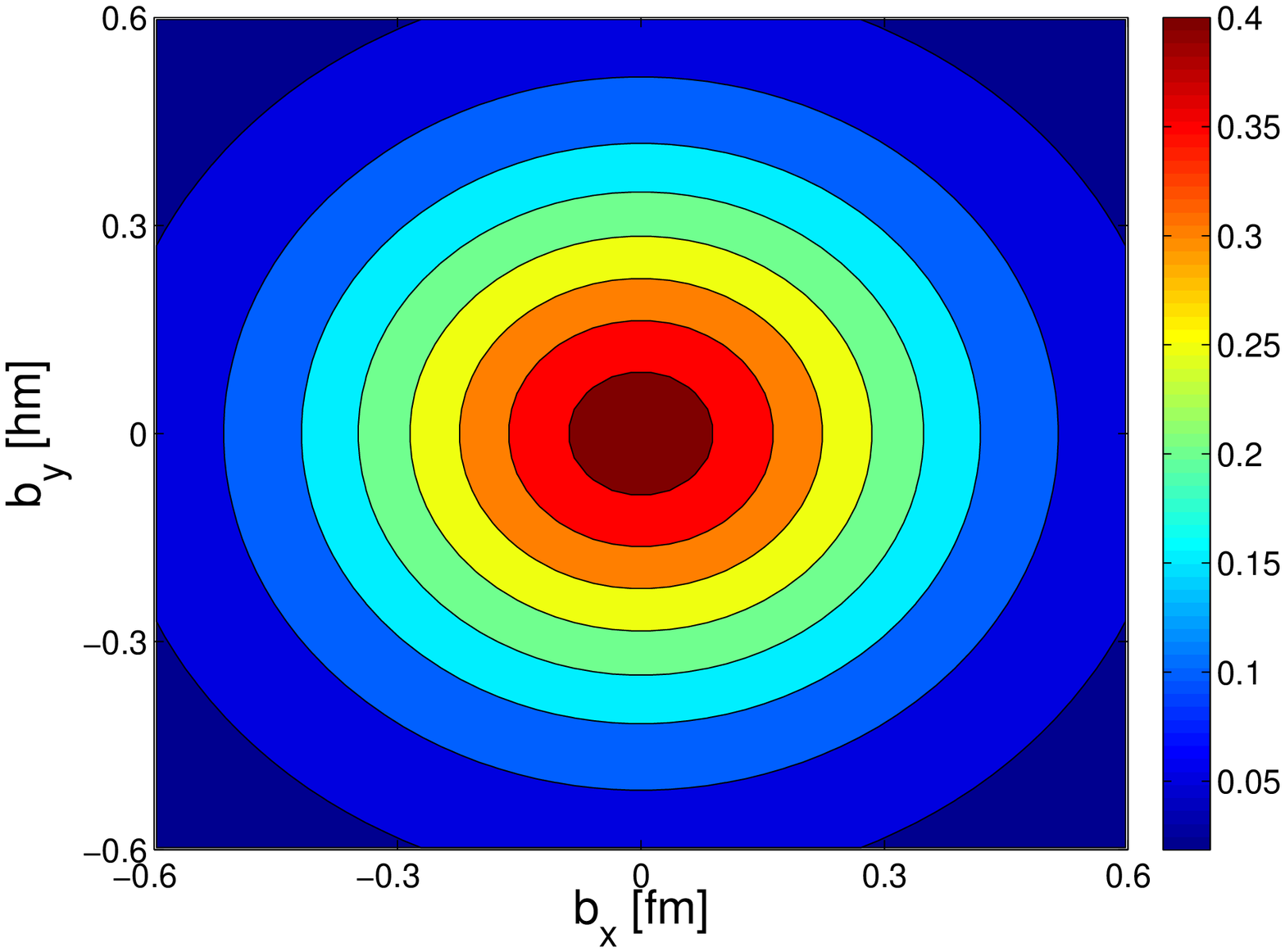}
\end{minipage}
\begin{minipage}[c]{0.98\textwidth}
\small{(c)}
\includegraphics[width=7cm,height=5.5cm,clip]{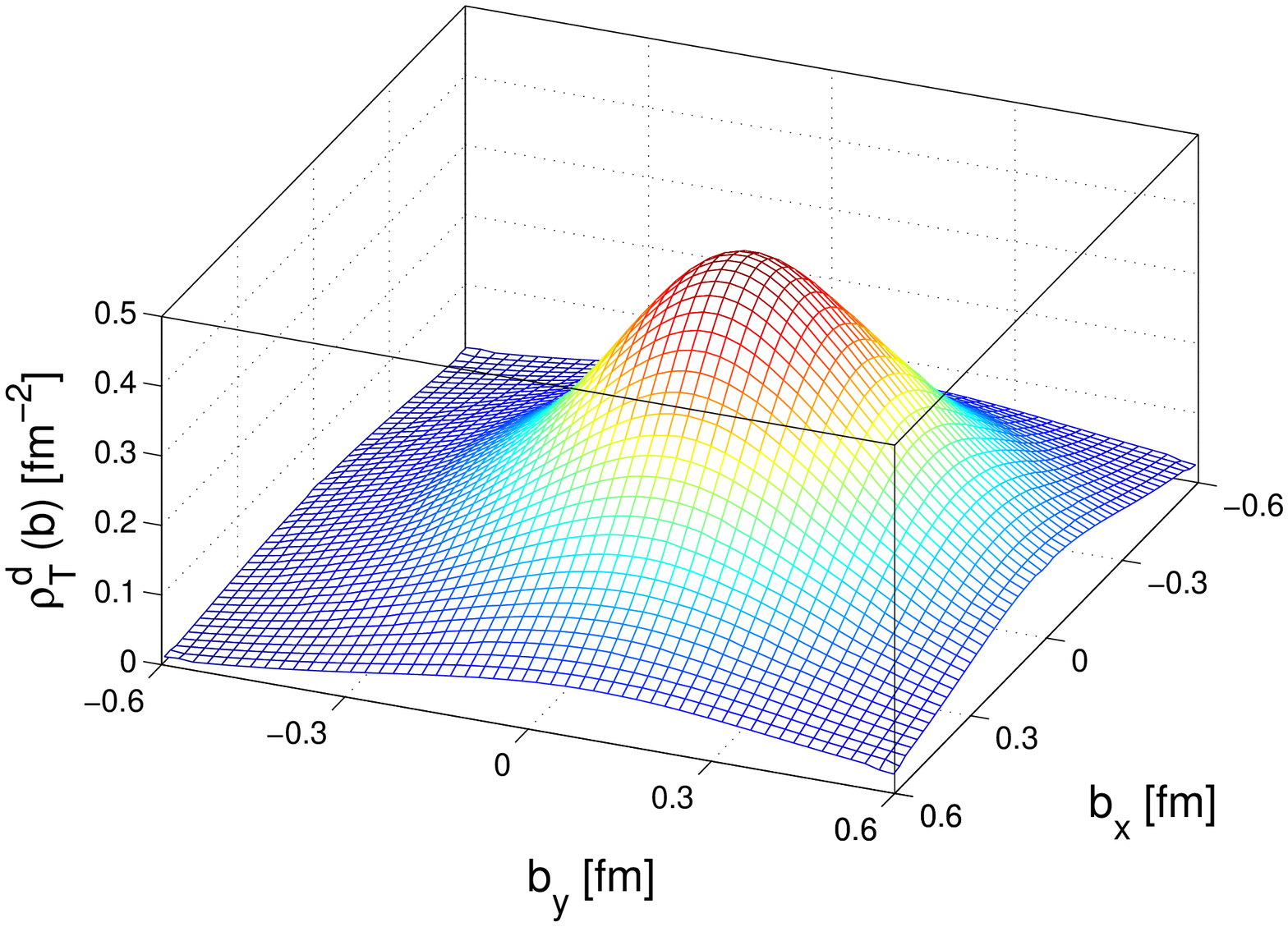}
\hspace{0.1cm}%
\small{(d)}\includegraphics[width=7cm,height=5.5cm,clip]{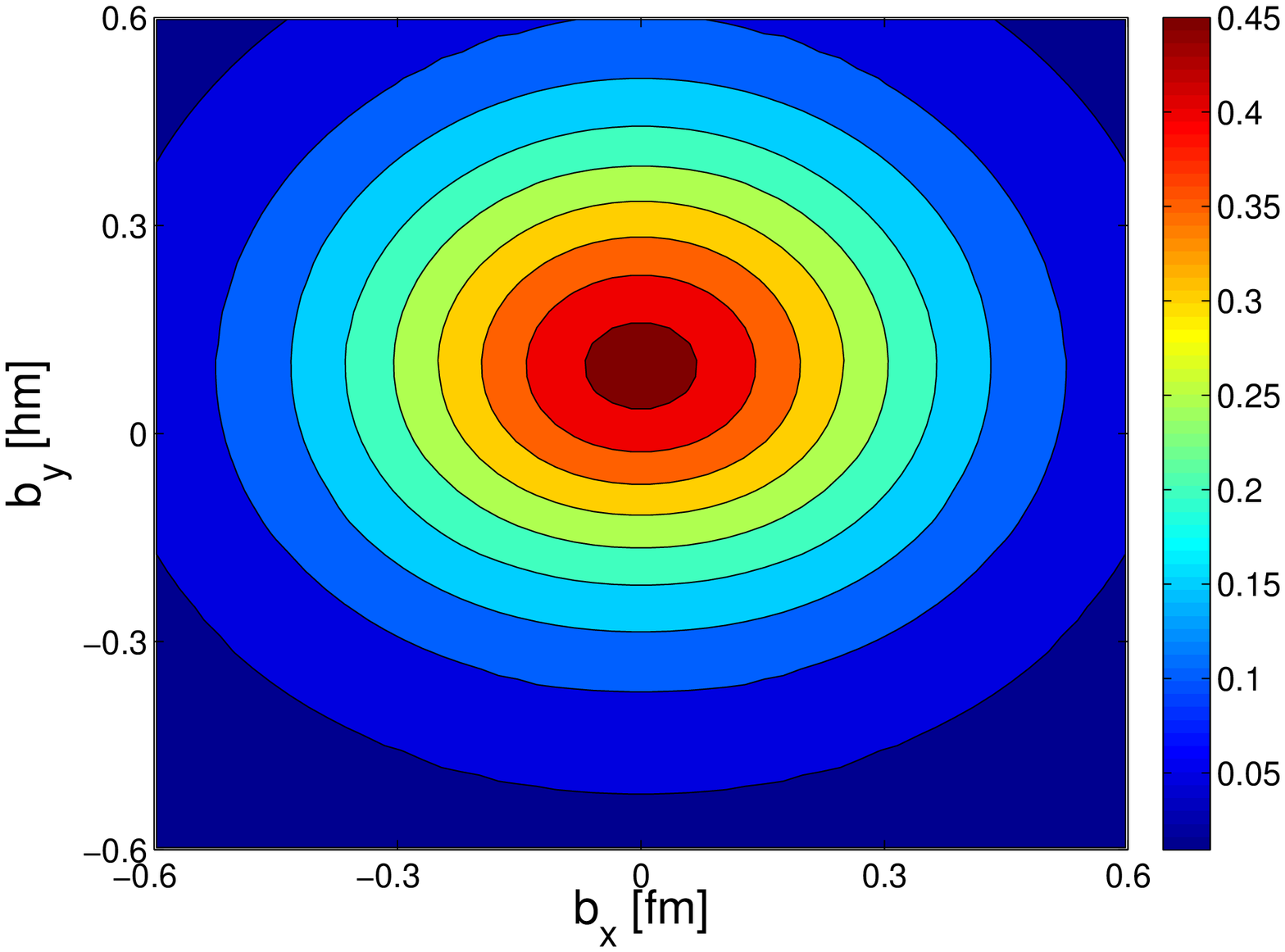}
\end{minipage}
\caption{\label{den_d}(Color online) The longitudinal momentum densities for the active $d$ quark in the transverse plane, upper panel for unpolarized nucleon, lower panel  for  nucleon polarized along $x$-direction. (b) and (d) are the top view of (a) and (c) respectively.}
\end{figure} 
\begin{figure}[htbp]
\begin{minipage}[c]{0.98\textwidth}
\small{(a)}
\includegraphics[width=7cm,height=5.5cm,clip]{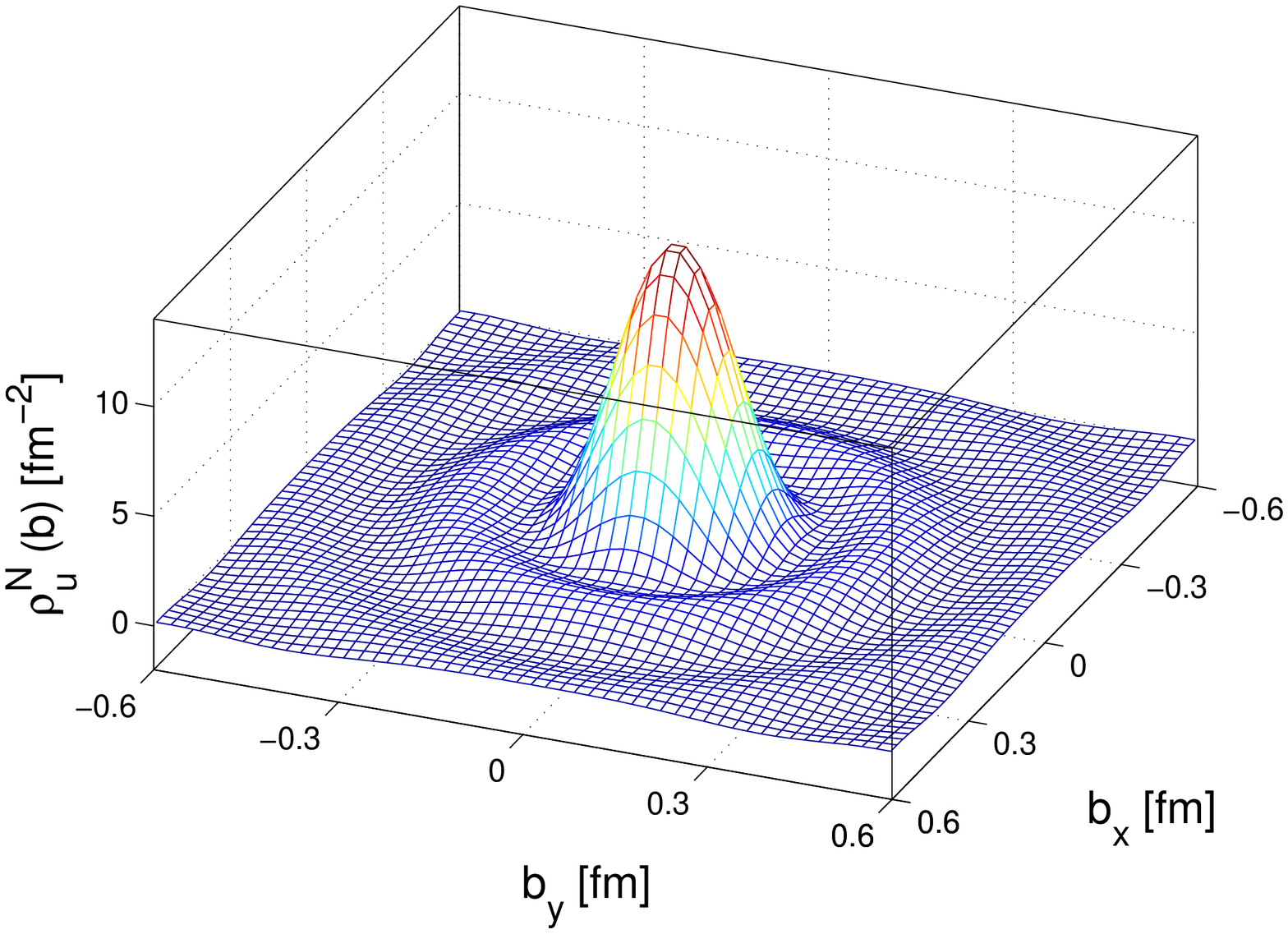}
\hspace{0.1cm}%
\small{(b)}\includegraphics[width=7cm,height=5.5cm,clip]{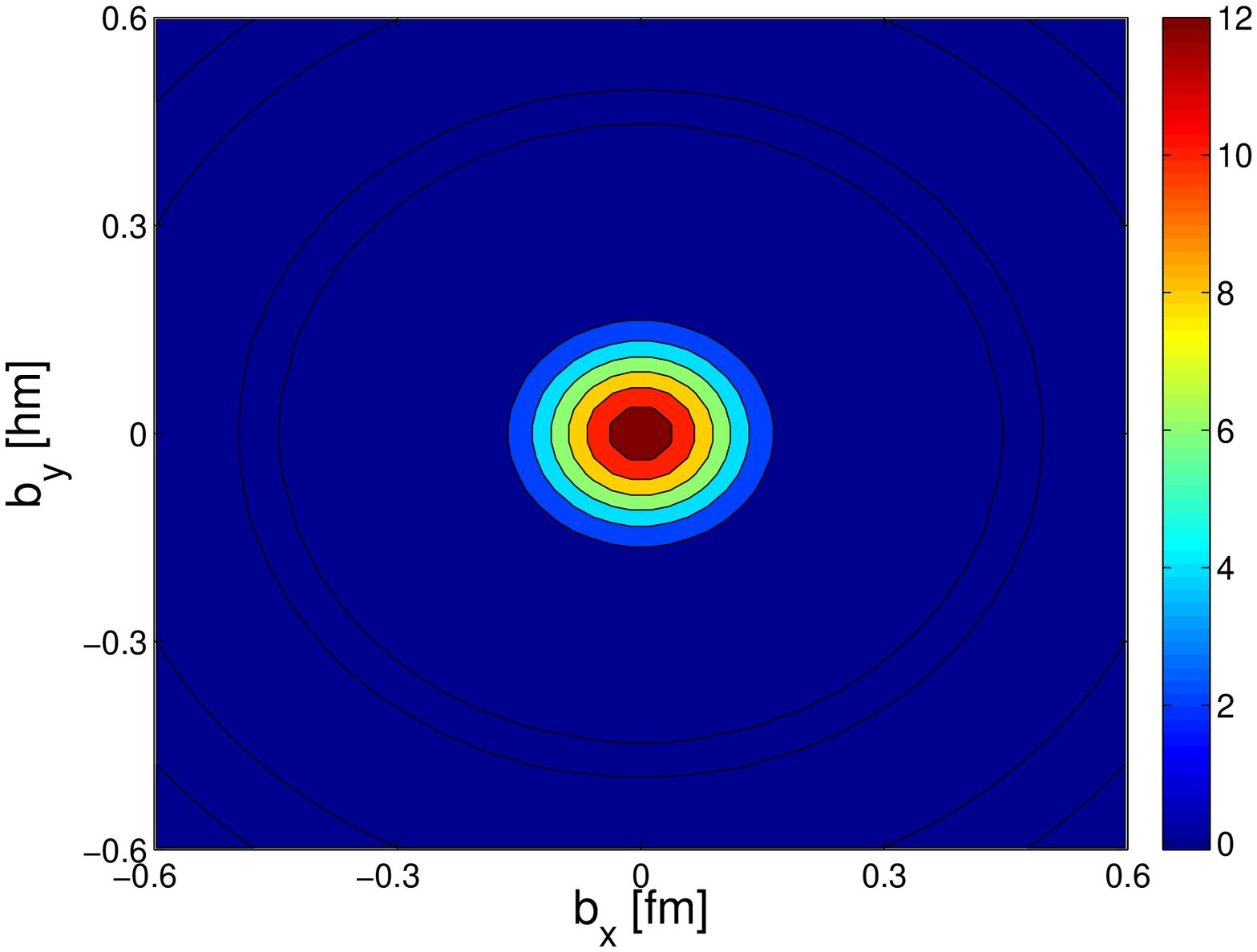}
\end{minipage}
\begin{minipage}[c]{0.98\textwidth}
\small{(c)}
\includegraphics[width=7cm,height=5.5cm,clip]{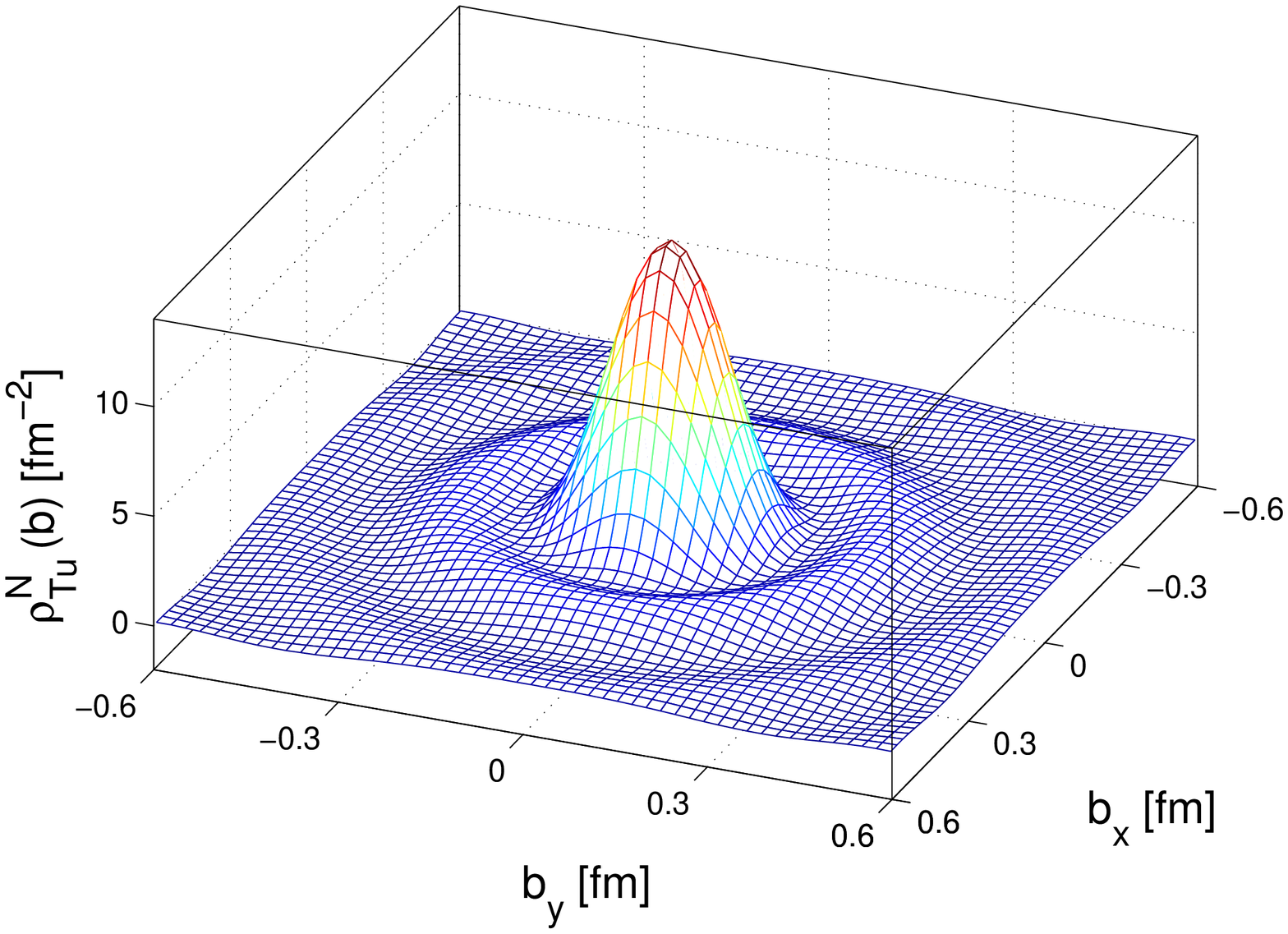}
\hspace{0.1cm}%
\small{(d)}\includegraphics[width=7cm,height=5.5cm,clip]{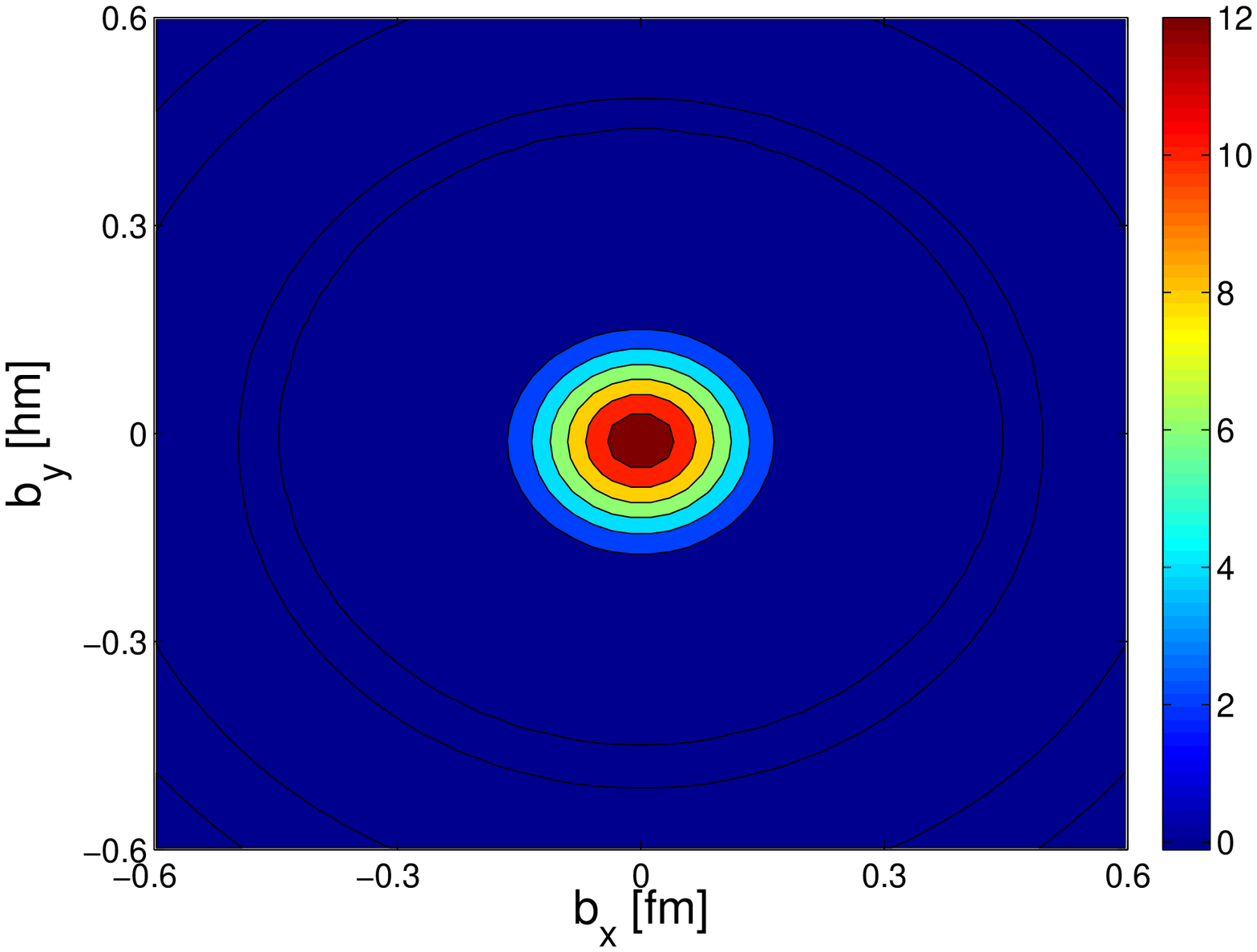}
\end{minipage}
\caption{\label{den_Nu}(Color online) The longitudinal momentum densities of nucleon in the transverse plane for the active quark $u$, upper panel for unpolarized nucleon, lower panel  for  nucleon polarized along $x$-direction. (b) and (d) are the top view of (a) and (c) respectively.}
\end{figure} 
\begin{figure}[htbp]
\begin{minipage}[c]{0.98\textwidth}
\small{(a)}
\includegraphics[width=7cm,height=5.5cm,clip]{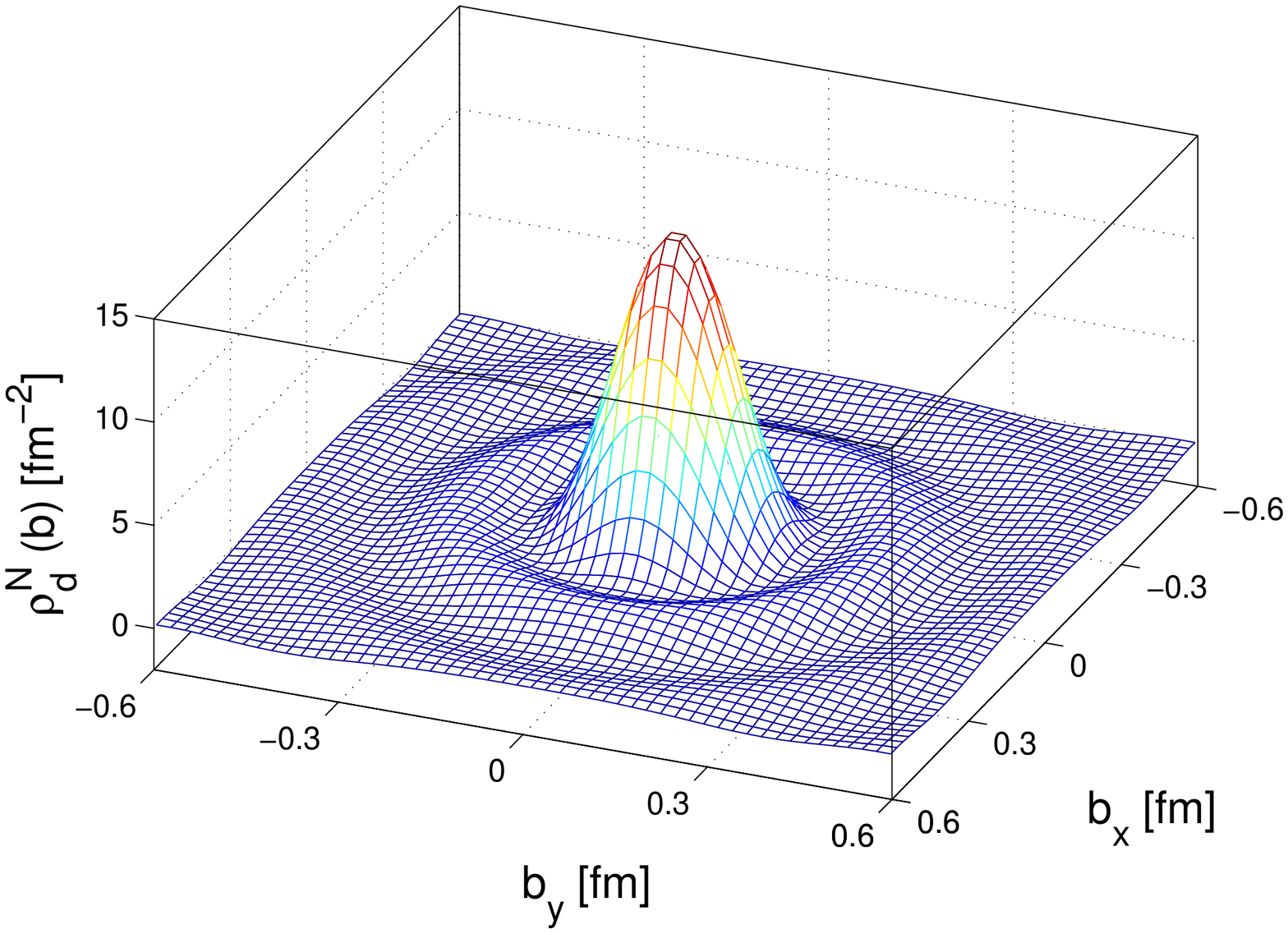}
\hspace{0.1cm}%
\small{(b)}\includegraphics[width=7cm,height=5.5cm,clip]{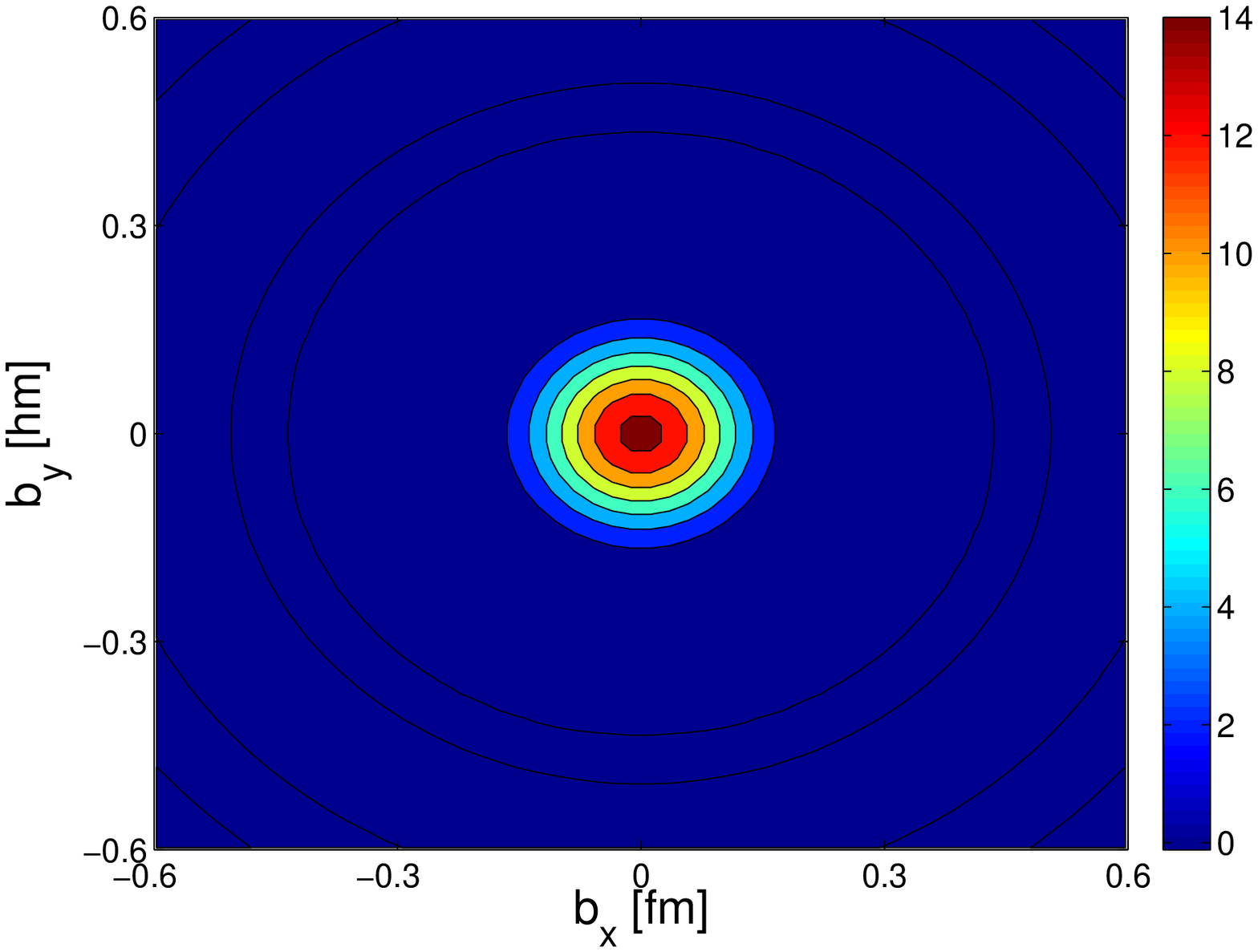}
\end{minipage}
\begin{minipage}[c]{0.98\textwidth}
\small{(c)}
\includegraphics[width=7cm,height=5.5cm,clip]{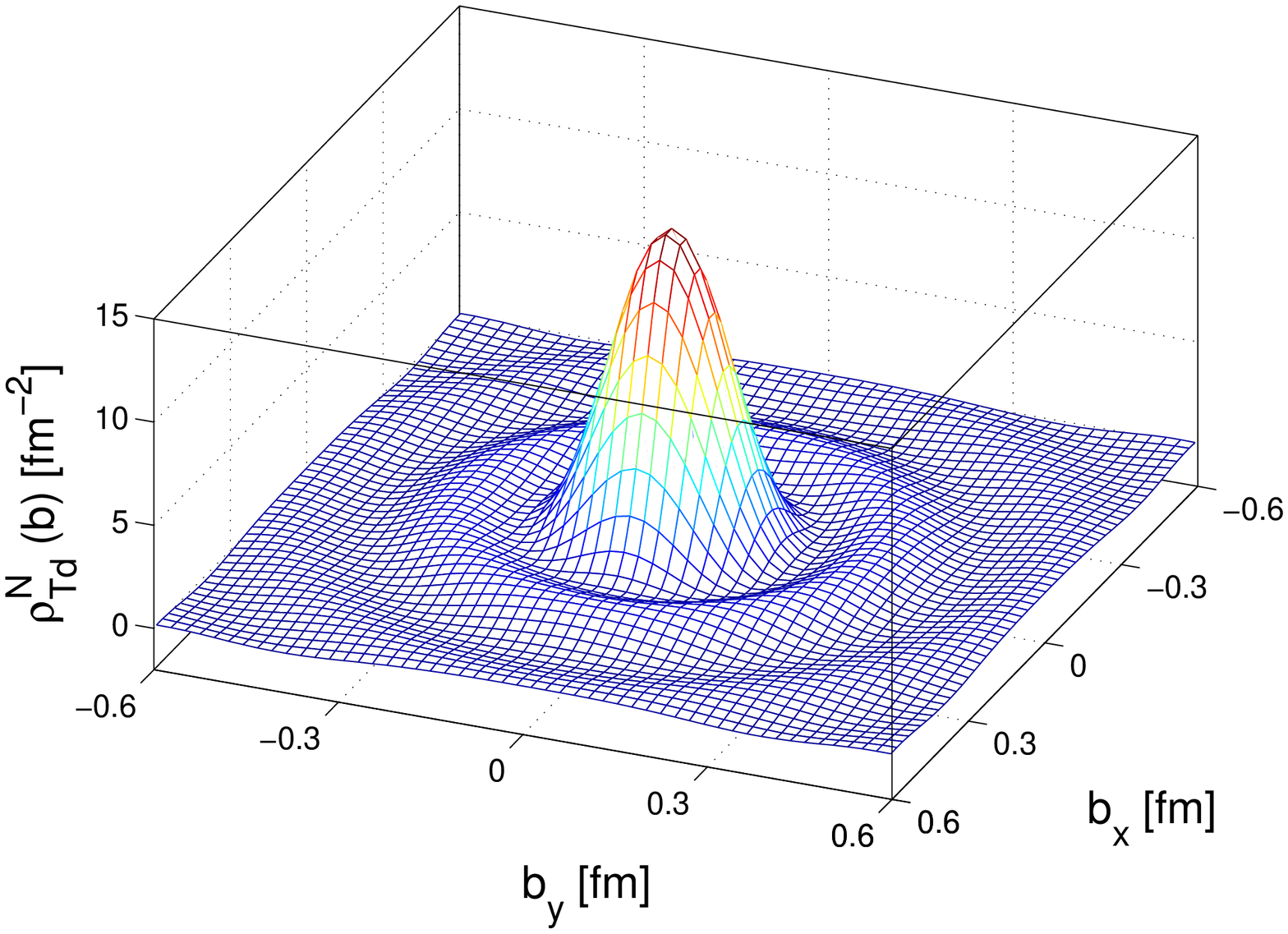}
\hspace{0.1cm}%
\small{(d)}\includegraphics[width=7cm,height=5.5cm,clip]{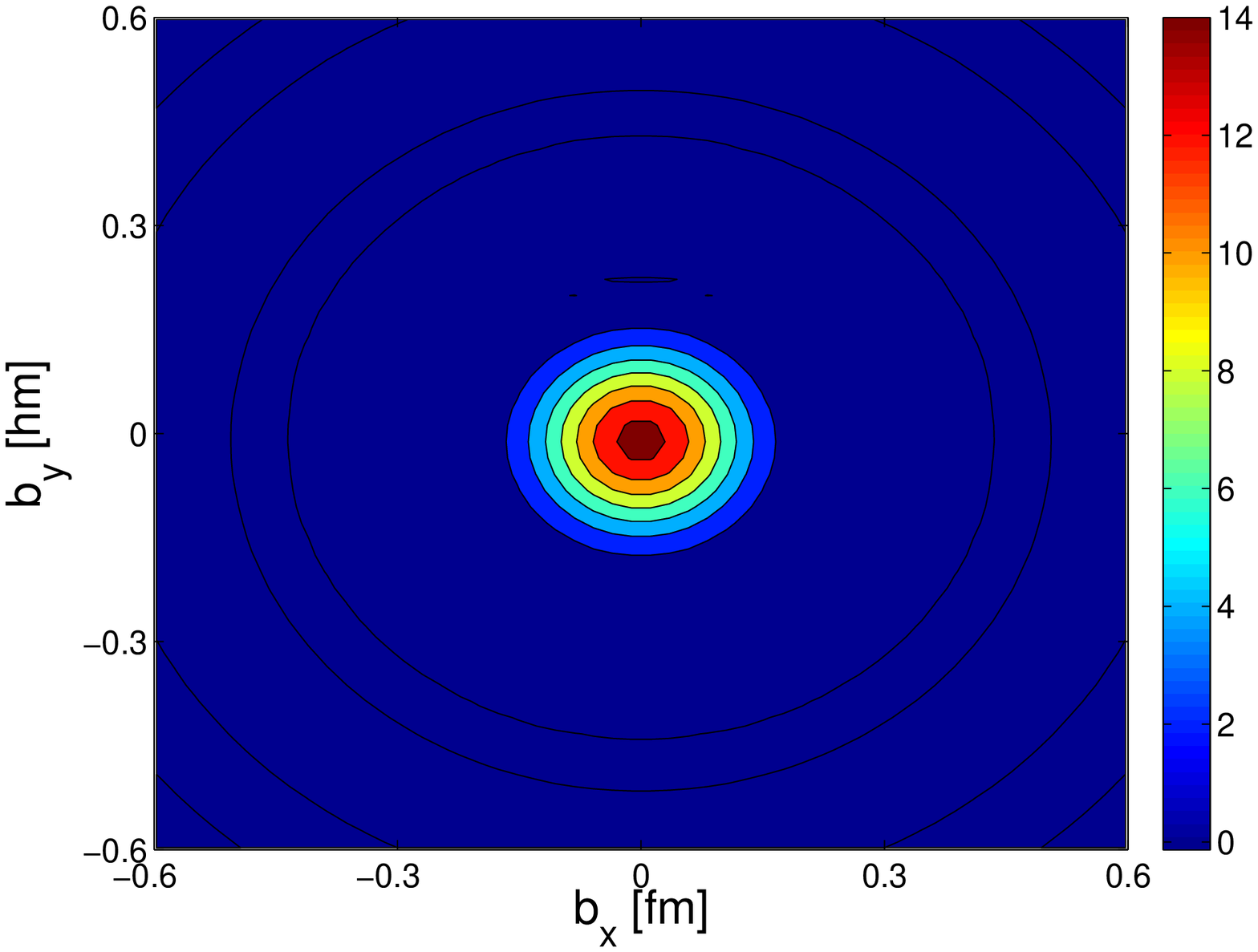}
\end{minipage}
\caption{\label{den_Nd}(Color online) The longitudinal momentum densities
of nucleon in the transverse plane for the active quark $d$, upper panel  for unpolarized nucleon, lower panel  for nucleon polarized along $x$-direction. (b) and (d) are the top view of (a) and (c) respectively.}
\end{figure} 
\begin{figure}[htbp]
\begin{minipage}[c]{0.98\textwidth}
\small{(a)}
\includegraphics[width=7cm,height=5.5cm,clip]{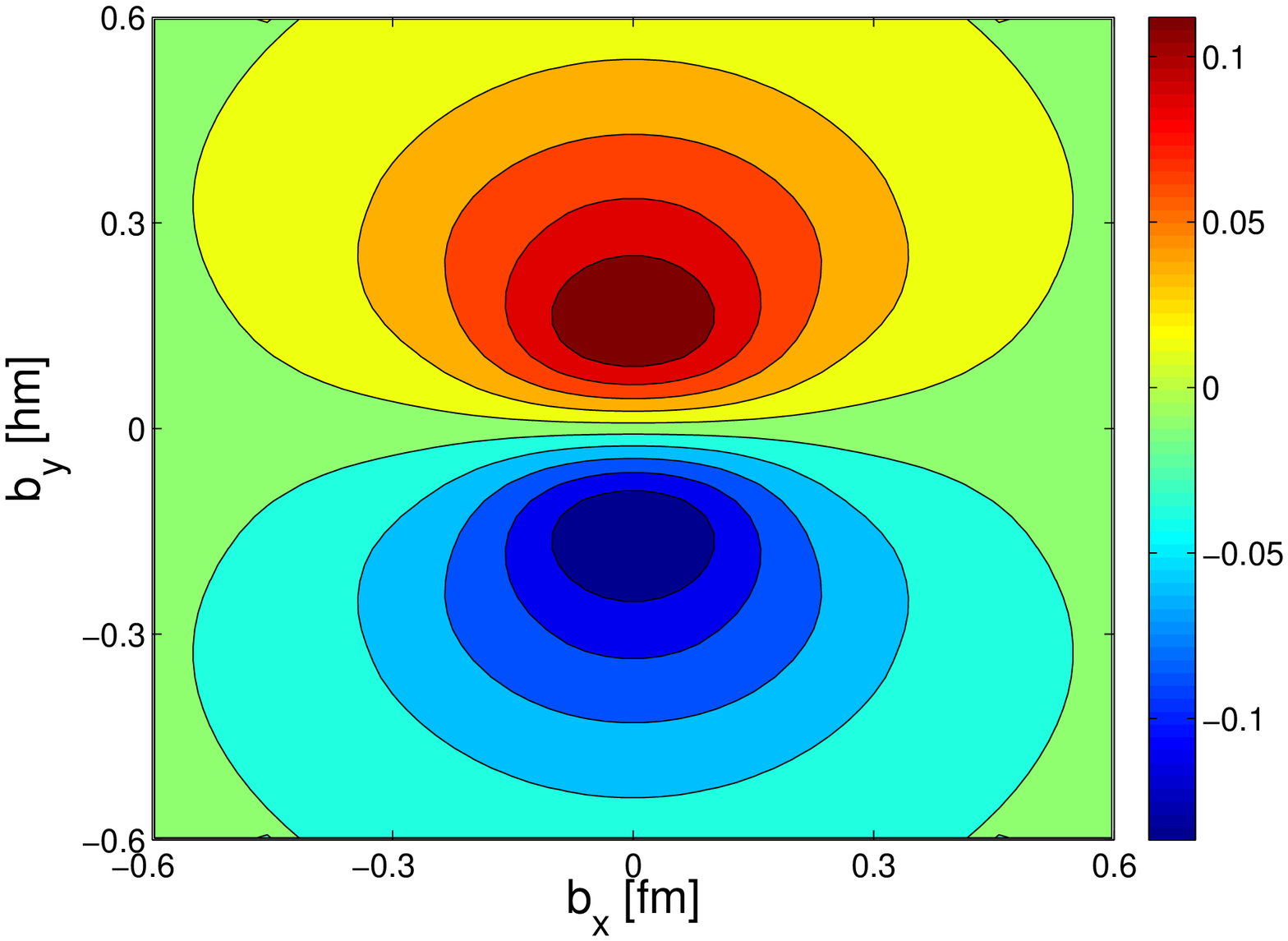}
\hspace{0.1cm}%
\small{(b)}\includegraphics[width=7cm,height=5.5cm,clip]{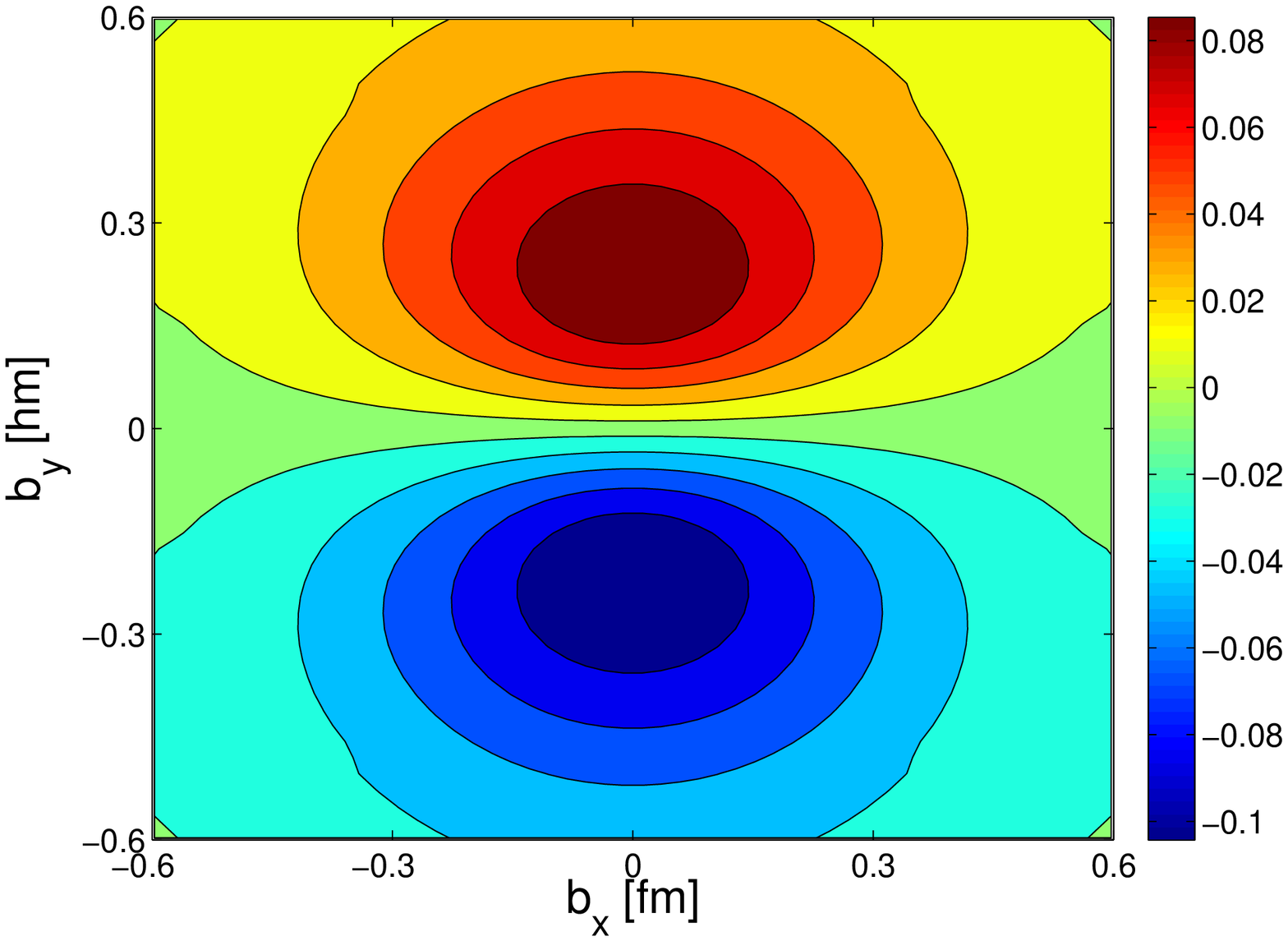}
\end{minipage}
\begin{minipage}[c]{0.98\textwidth}
\small{(c)}
\includegraphics[width=7cm,height=5.5cm,clip]{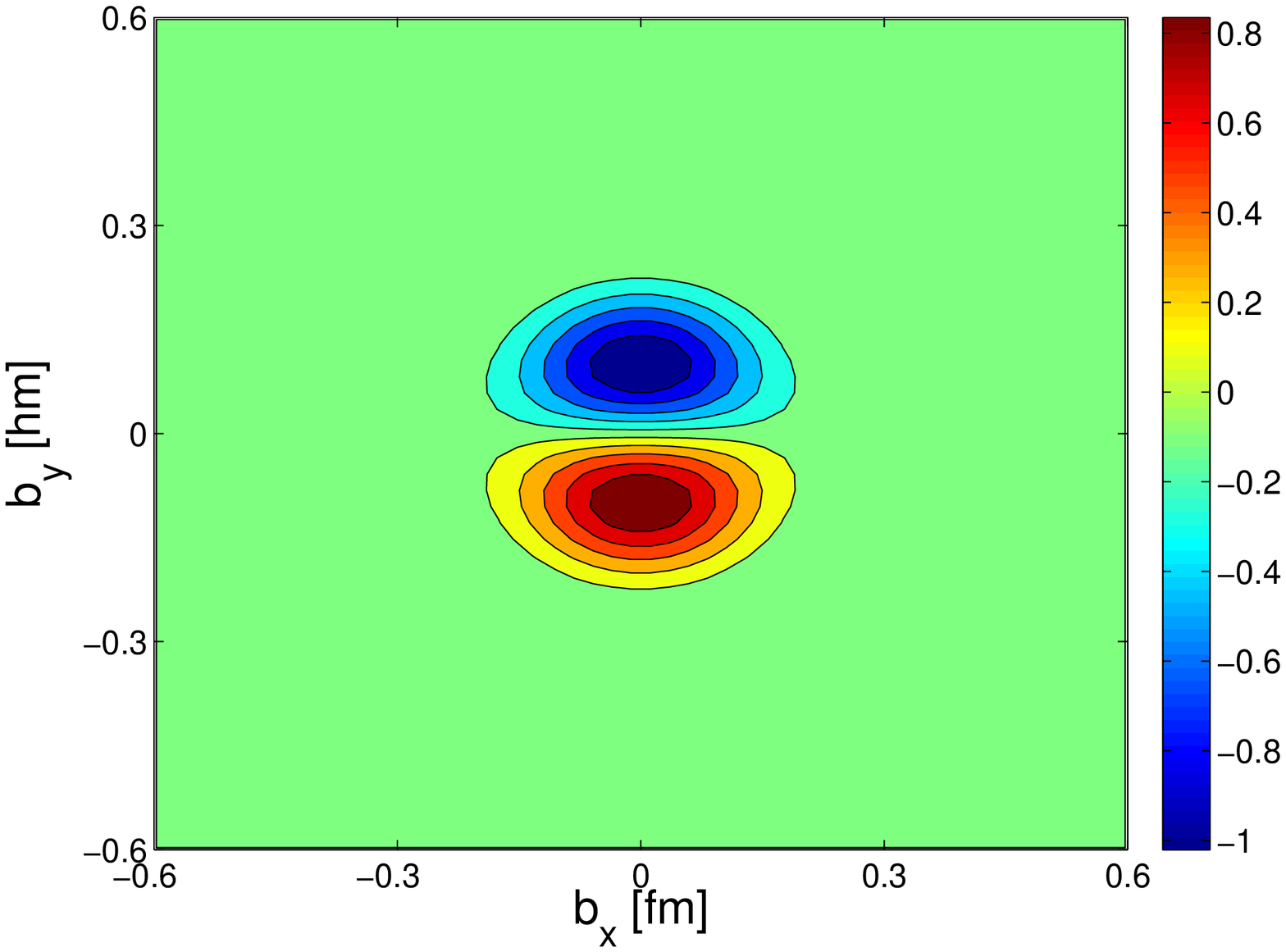}
\hspace{0.1cm}%
\small{(d)}\includegraphics[width=7cm,height=5.5cm,clip]{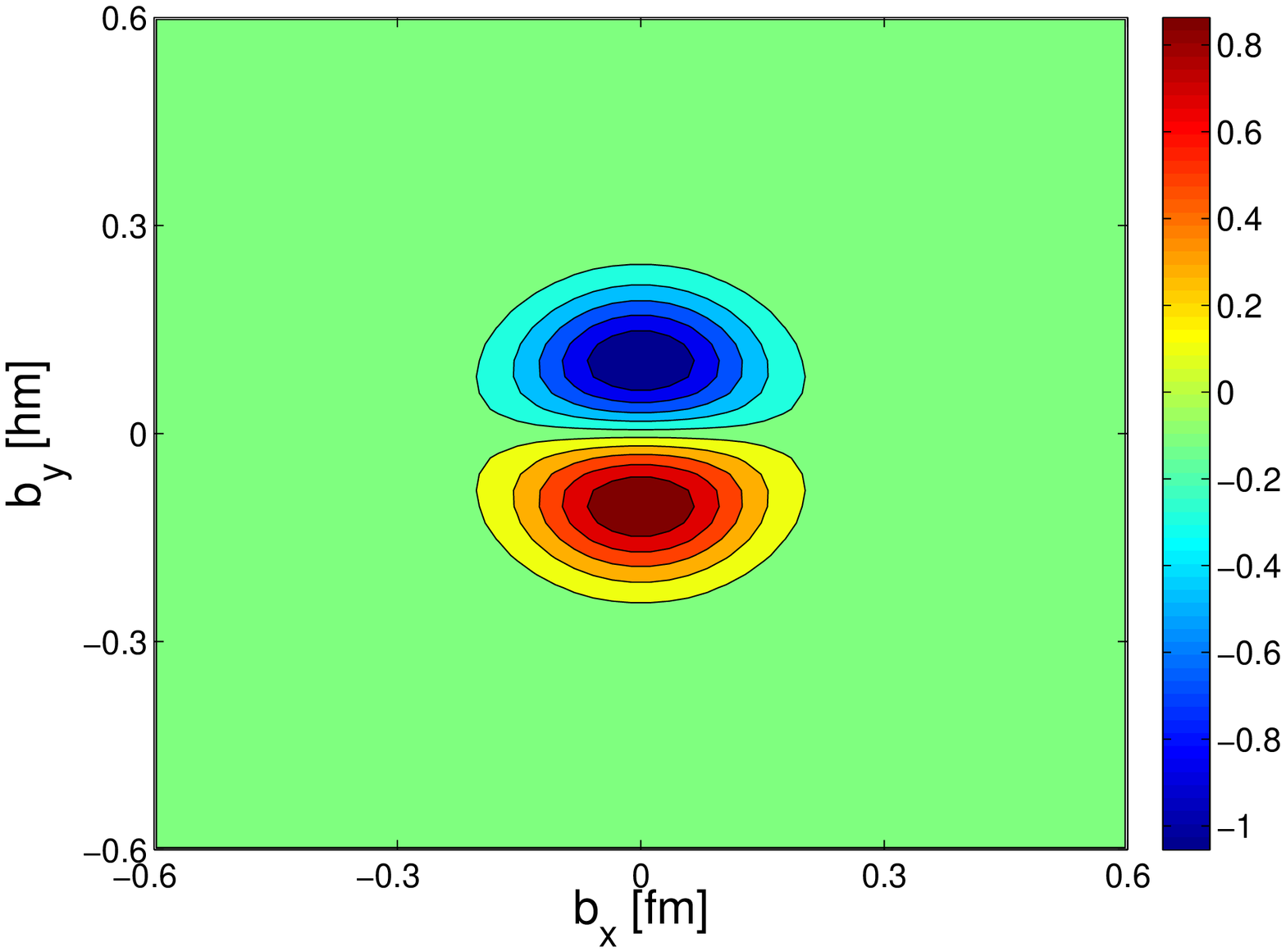}
\end{minipage}
\caption{\label{den_dipole}(Color online) The momentum density asymmetry $(\rho_T(b)-\rho(b))$ in the transverse plane for  a nucleon polarized in $x$-direction, (a) for $u$ quark (b) for $d$ quark (c) for nucleon when the active quark is $u$ and (d) for nucleon when the active quark is $d$.}
\end{figure} 
\section{Longitudinal momentum  density in transverse plane}\label{density}
According to the standard interpretation \cite{miller07,vande,weiss,CM3,selyugin} the charge and anomalous magnetization densities in the transverse plane can be identified with the two-dimensional Fourier transform(FT) of the electromagnetic form factors in the light-cone frame with $q^+=q^0+q^3=0$.  Similar to the electromagnetic densities, one can evaluate the gravitomagnetic density  in transverse plane by taking the FT of the gravitational form factor \cite{selyugin,abidin08}.
Since the longitudinal momentum is given by the $++$ component of the energy momentum tensor
\be P^+=\int dx^-d^2x^\perp T^{++},
\ee
it is possible to interpret the Fourier transform of the gravitational form factor $A(Q^2)$ as the longitudinal momentum density in the transverse plane \cite{abidin08}.
  For a unpolarized nucleon the momentum density  can be defined as 
\be
\rho(b)
=\int \frac{d^2q_{\perp}}{(2\pi)^2}A(Q^2)e^{iq_{\perp}.b_{\perp}}
=\int_0^\infty \frac{dQ}{2\pi}QJ_0(Qb)A(Q^2),
\ee
where $b=|{b_{\perp}}|$ represents the impact parameter and $J_0$ is the cylindrical Bessel function of order zero and $Q^2=q_{\perp}^2$. Under isospin symmetry, the momentum density is same for both proton and neutron.
Due to polarization, the density gets modified by a term which involves the spin flip form factor $B(Q^2)$. For transversely polarized nucleon, the momentum density is given by\cite{abidin08}
\be
\rho_T(b)=\rho(b)+\sin(\phi_b-\phi_s)\int_0^\infty\frac{dQ}{2\pi}\frac{Q^2}{2M_n}J_1(bQ)B(Q^2)\label{trans_pol},
\ee
where $M_n$ is the mass of nucleon. The transverse polarization of the nucleon is given by
$S_\perp=(\cos\phi_s \hat{x}+\sin\phi_s\hat{y})$ and the transverse impact parameter is denoted by $b_\perp=b(\cos\phi_b \hat{x} +\sin\phi_b\hat{y})$.  Without loss of generality, the polarization of the nucleon is chosen along $x$-axis ie., $\phi_s=0$. The second term in Eq.(\ref{trans_pol}), provides the deviation from circular symmetry of the unpolarized density.

Results for the momentum density $\rho (b)$ for the active quark $u$ for both unpolarized and the transversely polarized nucleon are shown in Fig.\ref{den_u}. Similar plots for the active quark $d$ are shown in Fig.\ref{den_d}. The plots show that the unpolarized densities are axially symmetric and have the peak at the center of the nucleon$(b=0)$. For the nucleon  polarized along $x$-direction, the peak of densities gets shifted towards positive $y$-direction and the densities no longer have the symmetry. It can also be noticed that the width of the density for $d$ quark is larger but the height of the peak is sufficiently small compare to $u$ quark. In Fig.\ref{den_Nu} and Fig.\ref{den_Nd}, we show the total nucleon $p^+$ densities (fermionic plus bosonic) for both the unpolarized and the transversely polarized nucleon for different active quark $u$ and $d$ respectively. The total angular-dependent part of the densities are small. So the shifting of the densities in $\rho_T(b)$ are also very small and it is practically invisible in Fig.\ref{den_Nu}(d) and Fig.\ref{den_Nd}(d). However, removing the
axially symmetric part of the density  from $\rho_T(b)$ i.e, if we look at $(\rho_T(b)-\rho(b))$, one can find that the total angular-dependent part of the density(i.e. distortion from the symmetry)  displays a dipole pattern (Fig.\ref{den_dipole}(c) and (d)). The angular-dependent part of the densities for active quarks $u$ and $d$ are shown in Fig.\ref{den_dipole}(a) and (b) respectively. Both plots show the dipole pattern but it is  broader for $d$ quark than $u$ quark.   Though the dipolar distortion for individual quark is quite large when both quark and diquark contributions are added together for the nucleon, the distortions become small irrespective of  struck quark flavor(Fig.\ref{den_dipole}(c) and (d)).

\section{Summary}\label{sum}
The main result of this work is to  show that the sum rules for the intrinsic spin for a transversely polarized proton involve the form factors $A_q$, $B_q$ and $\bar{C}_q$ in agreement with the claim in Ref.\cite{hari}. We demonstrated this in a recently proposed light front quark-diquark model where the LFWFs are modeled from the wave functions obtained from light front AdS/QCD. We have also shown explicit $Q^2$ behavior of the gravitational form factors in this model.  We have  evaluated the longitudinal momentum density ($p^+$ density) in the transverse  plane for  both unpolarized and polarized nucleon. For transversely polarized nucleon,   the asymmetries in the distributions for an individual  quark is quite large, but when the contributions from the quark and the bosonic diquark are considered, the overall asymmetries in the nucleon become small  but are shown to be dipolar in nature.

\appendix
\section{Matrix elements of $T^{\mu\nu}$}\label{appa}
\subsection{$T^{++}$ : up going to up matrix element}
\eq
T^{++}&=&\frac{i}{2}[\bar{\psi}\gamma^{+}(\overrightarrow{\partial}^{+}\psi)-\bar{\psi}\gamma^{+}\overleftarrow{\partial}^{+}\psi]+(\partial^{+}\phi)(\partial^{+}\phi) \nonumber \, \\  
&=&i[\psi^{\dagger}_{+}(\overrightarrow{\partial}^{+}\psi_+)-\psi^{\dagger}_+\overleftarrow{\partial}^{+}\psi_+]+(\partial^{+}\phi)(\partial^{+}\phi) \,  .
\en
\eq
\langle \Psi_{2p}^{\uparrow}\big(P^+,P^{\perp}=q\big)|T_{q}^{++}|\Psi_{2p}^{\uparrow}\big(P^+,P^{\perp}=0\big)\rangle
&=&\frac{\mathcal{F}_1^{q}(Q^2)}{I_1^q(0)},\nonumber\\
\langle \Psi_{2p}^{\uparrow}\big(P^+,P^{\perp}=q\big)|T_{b}^{++}|\Psi_{2p}^{\uparrow}\big(P^+,P^{\perp}=0\big)\rangle
&=&\frac{\mathcal{F}_1^{b}(Q^2)}{I_1^q(0)},
\en
where
\eq
\mathcal{F}_1^q(Q^2)&=&2(P^+)^2\int\frac{d^2k_{\perp}dx}{16\pi^3}  x~ 
\big[\psi_{+\frac{1}{2}}^{\uparrow *}(x,\vec{k'}_{\perp})\psi_{+\frac{1}{2}}^{\uparrow}(x,\vec{k}_{\perp})
+ \psi_{-\frac{1}{2}}^{\uparrow *}(x,\vec{k'}_{\perp})\psi_{-\frac{1}{2}}^{\uparrow}(x,\vec{k}_{\perp})\big] \, \nonumber\\
&=& 2(P^+)^2\int dx x \bigg[x^{2a_1}(1-x)^{2b_1+1}+\bigg(\frac{N_2}{N_1}\bigg)^2x^{2a_2-2}(1-x)^{2b_2+3}\frac{1}{M_n^2}\nonumber\\
&&\bigg(\frac{\kappa^2}{\log(1/x)}-\frac{Q^2}{4}\bigg)\bigg]\exp\bigg[-\frac{\log(1/x)}{\kappa^2}\frac{Q^2}{4}\bigg]\nonumber\\
&=& 2\bf(P^+)^2\bf{\mathcal{I}_{1q}}\,,
\en
\eq
\mathcal{F}_1^b(Q^2)&=&2(P^+)^2\int\frac{d^2k_{\perp}dx}{16\pi^3}  (1-x)~ 
\big[\psi_{+\frac{1}{2}}^{\uparrow *}(x,\vec{k''}_{\perp})\psi_{+\frac{1}{2}}^{\uparrow}(x,\vec{k}_{\perp})
+ \psi_{-\frac{1}{2}}^{\uparrow *}(x,\vec{k''}_{\perp})\psi_{-\frac{1}{2}}^{\uparrow}(x,\vec{k}_{\perp})\big] \, \nonumber\\
&=& 2(P^+)^2\int dx (1-x) \bigg[x^{2a_1}(1-x)^{2b_1+1}+\bigg(\frac{N_2}{N_1}\bigg)^2x^{2a_2-2}(1-x)^{2b_2+3}\frac{1}{M_n^2}\nonumber\\
&&\bigg(\frac{\kappa^2}{\log(1/x)}-\frac{x^2Q^2}{4(1-x)^2}\bigg)\bigg]\exp\bigg[-\frac{\log(1/x)}{\kappa^2}\frac{x^2Q^2}{4(1-x)^2}\bigg]\nonumber\\
&=& 2\bf(P^+)^2\bf{\mathcal{I}_{1b}}\,.
\en
\eq
\mathbf{\langle \Psi_{2p}^{\uparrow}\big(P^+,P^{\perp}=q\big)|T^{++}|\Psi_{2p}^{\uparrow}\big(P^+,P^{\perp}=0\big)\rangle
=2(P^+)^2\bigg[\frac{\mathcal{I}_{1q}(Q^2)}{I_1^q(0)}+\frac{\mathcal{I}_{1b}(Q^2)}{I_1^q(0)}\bigg]} \,.
\en
Using the matrix elements  Eq.(\ref{tensor}),
\eq
\bf{\langle \Psi_{2p}^{\uparrow}\big(P^+,P'^{\perp}=q\big)|T^{++}|\Psi_{2p}^{\uparrow}\big(P^+,P^{\perp}=0\big)\rangle}
=2(P^+)^2A(Q^2) \,.
\en
So,
\eq
\bf A(Q^2)=\bigg[\frac{\mathcal{I}_{1q}(Q^2)}{I_1^q(0)}+\frac{\mathcal{I}_{1b}(Q^2)}{I_1^q(0)}\bigg] \,.
\en
\subsection{$T^{++}$ : up going to down plus down going to up matrix elements}
\eq
\langle \Psi_{2p}^{\uparrow}\big(P'\big)|T_q^{++}|\Psi_{2p}^{\downarrow}\big(P\big)\rangle + \langle \Psi_{2p}^{\downarrow}\big(P'\big)|T_q^{++}|\Psi_{2p}^{\uparrow}\big(P\big)\rangle 
&=&\frac{\mathcal{F}_2^{q}(Q^2)}{I_2^q(0)},\nonumber\\
\langle \Psi_{2p}^{\uparrow}\big(P'\big)|T_b^{++}|\Psi_{2p}^{\downarrow}\big(P\big)\rangle + \langle \Psi_{2p}^{\downarrow}\big(P'\big)|T_b^{++}|\Psi_{2p}^{\uparrow}\big(P\big)\rangle 
&=&\frac{\mathcal{F}_2^{b}(Q^2)}{I_2^q(0)},
\en
where
\eq
\mathcal{F}_2^{q}(Q^2)
&=&2(P^+)^2\int\frac{d^2k_{\perp}dx}{16\pi^3}  x 
\bigg[\big\{\psi_{+\frac{1}{2}}^{\uparrow *}(x,\vec{k'}_{\perp})\psi_{+\frac{1}{2}}^{\downarrow}(x,\vec{k}_{\perp})
+ \psi_{-\frac{1}{2}}^{\uparrow *}(x,\vec{k'}_{\perp})\psi_{-\frac{1}{2}}^{\downarrow}(x,\vec{k}_{\perp})\big\}\nonumber\\
&&+\big\{\psi_{+\frac{1}{2}}^{\downarrow *}(x,\vec{k'}_{\perp})\psi_{+\frac{1}{2}}^{\uparrow}(x,\vec{k}_{\perp})
+ \psi_{-\frac{1}{2}}^{\downarrow *}(x,\vec{k'}_{\perp})\psi_{-\frac{1}{2}}^{\uparrow}(x,\vec{k}_{\perp})\big\}\bigg] \, \nonumber\\
&=&2(P^+)^2 (iq^2_{\perp})~ 2 \int dx~ \frac{N_2}{N_1} \frac{1}{M_n} x^{a_1+a_2}(1-x)^{b_1+b_2+2}\exp\bigg[-\frac{\log(1/x)}{\kappa^2}\frac{Q^2}{4}\bigg]\nonumber\\
&=&\bf 2(P^+)^2 (iq^2_{\perp})2\mathcal{I}_{2q} \,,
\en
\eq
\mathcal{F}_2^{b}(Q^2)
&=&2(P^+)^2\int\frac{d^2k_{\perp}dx}{16\pi^3}  (1-x) 
\bigg[\big\{\psi_{+\frac{1}{2}}^{\uparrow *}(x,\vec{k''}_{\perp})\psi_{+\frac{1}{2}}^{\downarrow}(x,\vec{k}_{\perp})
+ \psi_{-\frac{1}{2}}^{\uparrow *}(x,\vec{k''}_{\perp})\psi_{-\frac{1}{2}}^{\downarrow}(x,\vec{k}_{\perp})\big\}\nonumber\\
&&+\big\{\psi_{+\frac{1}{2}}^{\downarrow *}(x,\vec{k''}_{\perp})\psi_{+\frac{1}{2}}^{\uparrow}(x,\vec{k}_{\perp})
+ \psi_{-\frac{1}{2}}^{\downarrow *}(x,\vec{k''}_{\perp})\psi_{-\frac{1}{2}}^{\uparrow}(x,\vec{k}_{\perp})\big\}\bigg] \, \nonumber\\
&=&-2(P^+)^2 (iq^2_{\perp})~ 2 \int dx~ \frac{N_2}{N_1} \frac{1}{M_n} x^{a_1+a_2}(1-x)^{b_1+b_2+2}\exp\bigg[-\frac{\log(1/x)}{\kappa^2(1-x)^2}\frac{x^2Q^2}{4}\bigg]\nonumber\\
&=&-\bf 2(P^+)^2 (iq^2_{\perp})2\mathcal{I}_{2b} \,.
\en
\eq
\bf\langle \Psi_{2p}^{\uparrow}\big(P'\big)|T^{++}|\Psi_{2p}^{\downarrow}\big(P\big)\rangle &+& \bf\langle \Psi_{2p}^{\downarrow}\big(P'\big)|T^{++}|\Psi_{2p}^{\uparrow}\big(P\big)\rangle \nonumber\\
&=&\bf2(P^+)^2\bigg[\frac{2\mathcal{I}_{2q}(Q^2)}{I_2^q(0)}-\frac{2\mathcal{I}_{2b}(Q^2)}{I_2^q(0)}\bigg](iq^2_{\perp}) \,.
\en
Using the matrix elements from Eq.(\ref{tensor}),
\eq
\bf{\langle \Psi_{2p}^{\uparrow}\big(P'\big)|T^{++}|\Psi_{2p}^{\downarrow}\big(P\big)\rangle + \langle \Psi_{2p}^{\downarrow}\big(P'\big)|T^{++}|\Psi_{2p}^{\uparrow}\big(P\big)\rangle }=B(Q^2)\frac{2(P^+)^2}{M_n}(iq^2_{\perp}) \,.
\en
So,
\eq
\bf B(Q^2)=2M_n\bigg[\frac{\mathcal{I}_{2q}(Q^2)}{I_2^q(0)}-\frac{\mathcal{I}_{2b}(Q^2)}{I_2^q(0)}\bigg] \,.
\en
\subsection{$T^{+1}$ : up going to up matrix element}
\eq
T^{+1}&=&\frac{i}{2}[\bar{\psi}\gamma^{+}(\overrightarrow{\partial}^{1}\psi)-\bar{\psi}\gamma^{+}\overleftarrow{\partial}^{1}\psi]+(\partial^{+}\phi)(\partial^{1}\phi) \nonumber \, \\  
&=&i[\psi^{\dagger}_{+}(\overrightarrow{\partial}^{1}\psi_+)-\psi^{\dagger}_+\overleftarrow{\partial}^{1}\psi_+]+(\partial^{+}\phi)(\partial^{1}\phi) \,  .
\en
\eq
\langle \Psi_{2p}^{\uparrow}\big(P^+,P^{\perp}=q\big)|T_{q}^{+1}|\Psi_{2p}^{\uparrow}\big(P^+,P^{\perp}=0\big)\rangle
&=&\frac{\mathcal{F}_3^{q}(Q^2)}{I_1^q(0)},\nonumber\\
\langle \Psi_{2p}^{\uparrow}\big(P^+,P^{\perp}=q\big)|T_{b}^{+1}|\Psi_{2p}^{\uparrow}\big(P^+,P^{\perp}=0\big)\rangle
&=&\frac{\mathcal{F}_3^{b}(Q^2)}{I_1^q(0)},
\en
where
\eq
\mathcal{F}_3^q(Q^2)&=&2P^+\int\frac{d^2k_{\perp}dx}{16\pi^3}  (-k_1^{\perp})
\big[\psi_{+\frac{1}{2}}^{\uparrow *}(x,\vec{k'}_{\perp})\psi_{+\frac{1}{2}}^{\uparrow}(x,\vec{k}_{\perp})
+ \psi_{-\frac{1}{2}}^{\uparrow *}(x,\vec{k'}_{\perp})\psi_{-\frac{1}{2}}^{\uparrow}(x,\vec{k}_{\perp})\big] \, \nonumber\\
&=& P^+\int dx  \bigg[\bigg\{x^{2a_1}(1-x)^{2b_1+2}+\bigg(\frac{N_2}{N_1}\bigg)^2x^{2a_2-2}(1-x)^{2b_2+4}\frac{1}{M_n^2}\nonumber\\
&&\bigg(\frac{\kappa^2}{\log(1/x)}-\frac{Q^2}{4}\bigg)\bigg\}q^1_{\perp}+\bigg(\frac{N_2}{N_1}\bigg)^2x^{2a_2-2}(1-x)^{2b_2+4}\frac{1}{M_n^2}\frac{\kappa^2}{\log(1/x)}(iq^2_{\perp})\bigg]\nonumber\\
&&\times\exp\bigg[-\frac{\log(1/x)}{\kappa^2}\frac{Q^2}{4}\bigg]\nonumber\\
&=& \mathbf{ P^+(q^1_{\perp}\bf{\mathcal{I}_{3q}}+iq^2_{\perp}\bf{\mathcal{I}'_{3q}})}\,,
\en
\eq
\mathcal{F}_3^b(Q^2)&=&2P^+\int\frac{d^2k_{\perp}dx}{16\pi^3}  k_1^{\perp}
\big[\psi_{+\frac{1}{2}}^{\uparrow *}(x,\vec{k''}_{\perp})\psi_{+\frac{1}{2}}^{\uparrow}(x,\vec{k}_{\perp})
+ \psi_{-\frac{1}{2}}^{\uparrow *}(x,\vec{k''}_{\perp})\psi_{-\frac{1}{2}}^{\uparrow}(x,\vec{k}_{\perp})\big] \, \nonumber\\
&=& P^+\int dx  \bigg[\bigg\{x^{2a_1+1}(1-x)^{2b_1+1}+\bigg(\frac{N_2}{N_1}\bigg)^2x^{2a_2-1}(1-x)^{2b_2+3}\frac{1}{M_n^2}\nonumber\\
&&\bigg(\frac{\kappa^2}{\log(1/x)}-\frac{x^2Q^2}{4(1-x)^2}\bigg)\bigg\}q^1_{\perp}+\bigg(\frac{N_2}{N_1}\bigg)^2x^{2a_2-1}(1-x)^{2b_2+3}\frac{1}{M_n^2}\frac{\kappa^2}{\log(1/x)}(iq^2_{\perp})\bigg]\nonumber\\
&&\times\exp\bigg[-\frac{\log(1/x)}{\kappa^2(1-x)^2}\frac{x^2Q^2}{4}\bigg] \nonumber\\
&=& \mathbf{ P^+(q^1_{\perp}\bf{\mathcal{I}_{3b}}+iq^2_{\perp}\bf{\mathcal{I}'_{3b}})}\,,
\en

\eq \label{eq1}
\bf \langle \Psi_{2p}^{\uparrow}\big(P'\big)|T^{+1}|\Psi_{2p}^{\uparrow}\big(P\big)\rangle
=\frac{(\mathcal{I}_{3q}+\mathcal{I}_{3b})}{I_1^q(0)}P^+q^1_{\perp}+ \frac{(\mathcal{I}'_{3q}+\mathcal{I}'_{3b})}{I_1^q(0)}P^+(iq^2_{\perp})\,.
\en
Using the matrix elements from Eq.(\ref{tensor}),
\eq \label{eq2}
\bf{\langle \Psi_{2p}^{\uparrow}\big(P'\big)|T^{+1}|\Psi_{2p}^{\uparrow}\big(P\big)\rangle}
=A(Q^2)P^+(q^1_{\perp})+\frac{1}{2}(A(Q^2)+B(Q^2))P^+(-iq^2_{\perp}) \,.
\en
Comparing Eq. \ref{eq1} and Eq. \ref{eq2},
\eq 
\bf A(Q^2)&=&\bf\frac{(\mathcal{I}_{3q}+\mathcal{I}_{3b})}{I_1^q(0)} \nonumber \,, \\
\bf A(Q^2)+B(Q^2)&=&\bf-2\frac{(\mathcal{I}'_{3q}+\mathcal{I}'_{3b})}{I_1^q(0)} \, .
\en
\subsection{$T^{+1}$ : up going to down plus down going to up matrix elements}
\eq
\langle \Psi_{2p}^{\uparrow}\big(P'\big)|T_f^{+1}|\Psi_{2p}^{\downarrow}\big(P\big)\rangle + \langle \Psi_{2p}^{\downarrow}\big(P'\big)|T_f^{+1}|\Psi_{2p}^{\uparrow}\big(P\big)\rangle &=&\frac{\mathcal{F}_4^{q}(Q^2)}{I_2^q(0)},\nonumber\\
\langle \Psi_{2p}^{\uparrow}\big(P'\big)|T_b^{+1}|\Psi_{2p}^{\downarrow}\big(P\big)\rangle + \langle \Psi_{2p}^{\downarrow}\big(P'\big)|T_b^{+1}|\Psi_{2p}^{\uparrow}\big(P\big)\rangle &=&\frac{\mathcal{F}_4^{b}(Q^2)}{I_2^q(0)},
\en
where
\eq
\mathcal{F}_4^{q}(Q^2)
&=&2P^+\int\frac{d^2k_{\perp}dx}{16\pi^3}  (-k_1^{\perp})
\bigg[\big\{\psi_{+\frac{1}{2}}^{\uparrow *}(x,\vec{k'}_{\perp})\psi_{+\frac{1}{2}}^{\downarrow}(x,\vec{k}_{\perp})
+ \psi_{-\frac{1}{2}}^{\uparrow *}(x,\vec{k'}_{\perp})\psi_{-\frac{1}{2}}^{\downarrow}(x,\vec{k}_{\perp})\big\}\nonumber\\
&&+\big\{\psi_{+\frac{1}{2}}^{\downarrow *}(x,\vec{k'}_{\perp})\psi_{+\frac{1}{2}}^{\uparrow}(x,\vec{k}_{\perp})
+ \psi_{-\frac{1}{2}}^{\downarrow *}(x,\vec{k'}_{\perp})\psi_{-\frac{1}{2}}^{\uparrow}(x,\vec{k}_{\perp})\big\}\bigg] \, \nonumber\\
&=&2P^+ (iq^1_{\perp}q^2_{\perp})~ \int dx~ \frac{N_2}{N_1} \frac{1}{M_n} x^{a_1+a_2-1}(1-x)^{b_1+b_2+3}\exp\bigg[-\frac{\log(1/x)}{\kappa^2}\frac{Q^2}{4}\bigg]\nonumber\\
&=&\bf 2P^+ (iq^1_{\perp}q^2_{\perp})\mathcal{I}_{4q} \,,
\en
\eq
\mathcal{F}_4^{b}(Q^2)
&=&2P^+\int\frac{d^2k_{\perp}dx}{16\pi^3}   k_1^{\perp}
\bigg[\big\{\psi_{+\frac{1}{2}}^{\uparrow *}(x,\vec{k''}_{\perp})\psi_{+\frac{1}{2}}^{\downarrow}(x,\vec{k}_{\perp})
+ \psi_{-\frac{1}{2}}^{\uparrow *}(x,\vec{k''}_{\perp})\psi_{-\frac{1}{2}}^{\downarrow}(x,\vec{k}_{\perp})\big\}\nonumber\\
&&+\big\{\psi_{+\frac{1}{2}}^{\downarrow *}(x,\vec{k''}_{\perp})\psi_{+\frac{1}{2}}^{\uparrow}(x,\vec{k}_{\perp})
+ \psi_{-\frac{1}{2}}^{\downarrow *}(x,\vec{k''}_{\perp})\psi_{-\frac{1}{2}}^{\uparrow}(x,\vec{k}_{\perp})\big\}\bigg] \, \nonumber\\
&=&-2P^+ (iq^1_{\perp}q^2_{\perp}) \int dx~ \frac{N_2}{N_1} \frac{1}{M_n} x^{a_1+a_2+1}(1-x)^{b_1+b_2+1}\exp\bigg[-\frac{\log(1/x)}{\kappa^2(1-x)^2}\frac{x^2Q^2}{4}\bigg]\nonumber\\
&=&-\bf 2P^+ (iq^1_{\perp}q^2_{\perp})\mathcal{I}_{4b} \,.
\en
\eq
\bf{\langle \Psi_{2p}^{\uparrow}\big(P'\big)|T^{+1}|\Psi_{2p}^{\downarrow}\big(P\big)\rangle + \langle \Psi_{2p}^{\downarrow}\big(P'\big)|T^{+1}|\Psi_{2p}^{\uparrow}\big(P\big)\rangle }=\frac{2(\mathcal{I}_{4q}-\mathcal{I}_{4b})}{I_2^q(0)}P^+ (iq^1_{\perp}q^2_{\perp}) \,.
\en
Using the matrix elements from Eq.(\ref{tensor}),
\eq
\bf{\langle \Psi_{2p}^{\uparrow}\big(P'\big)|T^{+1}|\Psi_{2p}^{\downarrow}\big(P\big)\rangle + \langle \Psi_{2p}^{\downarrow}\big(P'\big)|T^{+1}|\Psi_{2p}^{\uparrow}\big(P\big)\rangle }=B(Q^2)\frac{P^+}{M_n}(iq^1_{\perp}q^2_{\perp}) \,.
\en
We keep only terms which are linear in $q^{\perp}$, then ignoring the term $q^1_{\perp}q^2_{\perp}$,
\eq
\bf{\langle \Psi_{2p}^{\uparrow}\big(P'\big)|T^{+1}|\Psi_{2p}^{\downarrow}\big(P\big)\rangle + \langle \Psi_{2p}^{\downarrow}\big(P'\big)|T^{+1}|\Psi_{2p}^{\uparrow}\big(P\big)\rangle }=0 \,.
\en
\subsection{$T^{+2}$ : up going to up matrix element}
\eq
T^{+2}&=&\frac{i}{2}[\bar{\psi}\gamma^{+}(\overrightarrow{\partial}^{2}\psi)-\bar{\psi}\gamma^{+}\overleftarrow{\partial}^{2}\psi]+(\partial^{+}\phi)(\partial^{2}\phi) \nonumber \, \\  
&=&i[\psi^{\dagger}_{+}(\overrightarrow{\partial}^{2}\psi_+)-\psi^{\dagger}_+\overleftarrow{\partial}^{2}\psi_+]+(\partial^{+}\phi)(\partial^{2}\phi) \,  .
\en
\eq
\langle \Psi_{2p}^{\uparrow}\big(P^+,P^{\perp}=q\big)|T_{q}^{+2}|\Psi_{2p}^{\uparrow}\big(P^+,P^{\perp}=0\big)\rangle
&=&\frac{\mathcal{F}_5^{q}(Q^2)}{I_1^q(0)},\nonumber\\
\langle \Psi_{2p}^{\uparrow}\big(P^+,P^{\perp}=q\big)|T_{b}^{+2}|\Psi_{2p}^{\uparrow}\big(P^+,P^{\perp}=0\big)\rangle
&=&\frac{\mathcal{F}_5^{b}(Q^2)}{I_1^q(0)},
\en
where
\eq
\mathcal{F}_5^q(Q^2)&=&2P^+\int\frac{d^2k_{\perp}dx}{16\pi^3}  (-k_2^{\perp})
\big[\psi_{+\frac{1}{2}}^{\uparrow *}(x,\vec{k'}_{\perp})\psi_{+\frac{1}{2}}^{\uparrow}(x,\vec{k}_{\perp})
+ \psi_{-\frac{1}{2}}^{\uparrow *}(x,\vec{k'}_{\perp})\psi_{-\frac{1}{2}}^{\uparrow}(x,\vec{k}_{\perp})\big] \, \nonumber\\
&=& P^+\int dx  \bigg[\bigg\{x^{2a_1}(1-x)^{2b_1+2}+\bigg(\frac{N_2}{N_1}\bigg)^2x^{2a_2-2}(1-x)^{2b_2+4}\frac{1}{M_n^2}\nonumber\\
&&\bigg(\frac{\kappa^2}{\log(1/x)}-\frac{Q^2}{4}\bigg)\bigg\}q^2_{\perp}+\bigg(\frac{N_2}{N_1}\bigg)^2x^{2a_2-2}(1-x)^{2b_2+4}\frac{1}{M_n^2}\frac{\kappa^2}{\log(1/x)}(-iq^1_{\perp})\bigg]\nonumber\\
&&\times\exp\bigg[-\frac{\log(1/x)}{\kappa^2}\frac{Q^2}{4}\bigg] \nonumber\\
&=& \mathbf{ P^+(q^2_{\perp}\bf{\mathcal{I}_{5q}}-iq^1_{\perp}\bf{\mathcal{I}'_{5q}})}\,,
\en
\eq
\mathcal{F}_5^b(Q^2)&=&2P^+\int\frac{d^2k_{\perp}dx}{16\pi^3}  k_2^{\perp}
\big[\psi_{+\frac{1}{2}}^{\uparrow *}(x,\vec{k''}_{\perp})\psi_{+\frac{1}{2}}^{\uparrow}(x,\vec{k}_{\perp})
+ \psi_{-\frac{1}{2}}^{\uparrow *}(x,\vec{k''}_{\perp})\psi_{-\frac{1}{2}}^{\uparrow}(x,\vec{k}_{\perp})\big] \, \nonumber\\
&=& P^+\int dx  \bigg[\bigg\{x^{2a_1+1}(1-x)^{2b_1+1}+\bigg(\frac{N_2}{N_1}\bigg)^2x^{2a_2-1}(1-x)^{2b_2+3}\frac{1}{M_n^2}\nonumber\\
&&\bigg(\frac{\kappa^2}{\log(1/x)}-\frac{x^2Q^2}{4(1-x)^2}\bigg)\bigg\}q^2_{\perp}+\bigg(\frac{N_2}{N_1}\bigg)^2x^{2a_2-1}(1-x)^{2b_2+3}\frac{1}{M_n^2}\frac{\kappa^2}{\log(1/x)}(-iq^1_{\perp})\bigg]\nonumber\\
&&\times\exp\bigg[-\frac{\log(1/x)}{\kappa^2(1-x)^2}\frac{x^2Q^2}{4}\bigg] \nonumber\\
&=& \mathbf{ P^+(q^2_{\perp}\bf{\mathcal{I}_{5b}}-iq^1_{\perp}\bf{\mathcal{I}'_{5b}})}\,,
\en
\eq \label{eqT+21}
\bf \langle \Psi_{2p}^{\uparrow}\big(P'\big)|T^{+2}|\Psi_{2p}^{\uparrow}\big(P\big)\rangle
=\frac{(\mathcal{I}_{5q}+\mathcal{I}_{5b})}{I_1^q(0)}P^+q^2_{\perp}- \frac{(\mathcal{I}'_{5q}+\mathcal{I}'_{5b})}{I_1^q(0)}P^+(iq^1_{\perp})\,.
\en
Using the matrix elements from Eq.(\ref{tensor}),
\eq \label{eqT+22}
\bf{\langle \Psi_{2p}^{\uparrow}\big(P'\big)|T^{+2}|\Psi_{2p}^{\uparrow}\big(P\big)\rangle}
=A(Q^2)P^+q^2_{\perp}+\frac{1}{2}(A(Q^2)+B(Q^2))P^+(iq^1_{\perp}) \,.
\en
Comparing Eq. \ref{eqT+21} and Eq. \ref{eqT+22},
\eq 
\bf A(Q^2)&=&\bf\frac{(\mathcal{I}_{5q}+\mathcal{I}_{5b})}{I_1^q(0)} \nonumber \,, \\
\bf A(Q^2)+B(Q^2)&=&\bf-2\frac{(\mathcal{I}'_{5q}+\mathcal{I}'_{5b})}{I_1^q(0)} \, .
\en
\subsection{$T^{+2}$ : up going to down plus down going to up matrix elements}
\eq
\langle \Psi_{2p}^{\uparrow}\big(P'\big)|T_f^{+2}|\Psi_{2p}^{\downarrow}\big(P\big)\rangle + \langle \Psi_{2p}^{\downarrow}\big(P'\big)|T_f^{+2}|\Psi_{2p}^{\uparrow}\big(P\big)\rangle &=&\frac{\mathcal{F}_6^{q}(Q^2)}{I_2^q(0)},\nonumber\\
\langle \Psi_{2p}^{\uparrow}\big(P'\big)|T_b^{+2}|\Psi_{2p}^{\downarrow}\big(P\big)\rangle + \langle \Psi_{2p}^{\downarrow}\big(P'\big)|T_b^{+2}|\Psi_{2p}^{\uparrow}\big(P\big)\rangle &=&\frac{\mathcal{F}_6^{b}(Q^2)}{I_2^q(0)},
\en
where
\eq
\mathcal{F}_6^{q}(Q^2)
&=&2P^+\int\frac{d^2k_{\perp}dx}{16\pi^3}  (-k_2^{\perp})
\bigg[\big\{\psi_{+\frac{1}{2}}^{\uparrow *}(x,\vec{k'}_{\perp})\psi_{+\frac{1}{2}}^{\downarrow}(x,\vec{k}_{\perp})
+ \psi_{-\frac{1}{2}}^{\uparrow *}(x,\vec{k'}_{\perp})\psi_{-\frac{1}{2}}^{\downarrow}(x,\vec{k}_{\perp})\big\}\nonumber\\
&&+\big\{\psi_{+\frac{1}{2}}^{\downarrow *}(x,\vec{k'}_{\perp})\psi_{+\frac{1}{2}}^{\uparrow}(x,\vec{k}_{\perp})
+ \psi_{-\frac{1}{2}}^{\downarrow *}(x,\vec{k'}_{\perp})\psi_{-\frac{1}{2}}^{\uparrow}(x,\vec{k}_{\perp})\big\}\bigg] \, \nonumber\\
&=&2P^+ i(q^2_{\perp})^2~ \int dx~ \frac{N_2}{N_1} \frac{1}{M_n} x^{a_1+a_2-1}(1-x)^{b_1+b_2+3}\exp\bigg[-\frac{\log(1/x)}{\kappa^2}\frac{Q^2}{4}\bigg]\nonumber\\
&=&\bf 2P^+ i(q^2_{\perp})^2\mathcal{I}_{6q} \,,
\en
\eq
\mathcal{F}_6^{b}(Q^2)
&=&2P^+\int\frac{d^2k_{\perp}dx}{16\pi^3}   k_2^{\perp}
\bigg[\big\{\psi_{+\frac{1}{2}}^{\uparrow *}(x,\vec{k''}_{\perp})\psi_{+\frac{1}{2}}^{\downarrow}(x,\vec{k}_{\perp})
+ \psi_{-\frac{1}{2}}^{\uparrow *}(x,\vec{k''}_{\perp})\psi_{-\frac{1}{2}}^{\downarrow}(x,\vec{k}_{\perp})\big\}\nonumber\\
&&+\big\{\psi_{+\frac{1}{2}}^{\downarrow *}(x,\vec{k''}_{\perp})\psi_{+\frac{1}{2}}^{\uparrow}(x,\vec{k}_{\perp})
+ \psi_{-\frac{1}{2}}^{\downarrow *}(x,\vec{k''}_{\perp})\psi_{-\frac{1}{2}}^{\uparrow}(x,\vec{k}_{\perp})\big\}\bigg] \, \nonumber\\
&=&-2P^+ i(q^2_{\perp})^2 \int dx~ \frac{N_2}{N_1} \frac{1}{M_n} x^{a_1+a_2+1}(1-x)^{b_1+b_2+1}\exp\bigg[-\frac{\log(1/x)}{\kappa^2(1-x)^2}\frac{x^2Q^2}{4}\bigg]\nonumber\\
&=&\bf -2P^+ i(q^2_{\perp})^2\mathcal{I}_{6b} \,.
\en
\eq
\bf{\langle \Psi_{2p}^{\uparrow}\big(P'\big)|T^{+2}|\Psi_{2p}^{\downarrow}\big(P\big)\rangle + \langle \Psi_{2p}^{\downarrow}\big(P'\big)|T^{+2}|\Psi_{2p}^{\uparrow}\big(P\big)\rangle }=\frac{2(\mathcal{I}_{6q}-\mathcal{I}_{6b})}{I_2^q(0)}P^+ i(q^2_{\perp})^2 \,.
\en
Using the matrix elements from Eq.(\ref{tensor}),
\eq
\bf{\langle \Psi_{2p}^{\uparrow}\big(P'\big)|T^{+2}|\Psi_{2p}^{\downarrow}\big(P\big)\rangle + \langle \Psi_{2p}^{\downarrow}\big(P'\big)|T^{+2}|\Psi_{2p}^{\uparrow}\big(P\big)\rangle }=B(Q^2)\frac{P^+}{M_n}i(q^2_{\perp})^2 \,.
\en
Ignoring the term $(q^2_{\perp})^2$,
\eq
\bf{\langle \Psi_{2p}^{\uparrow}\big(P'\big)|T^{+2}|\Psi_{2p}^{\downarrow}\big(P\big)\rangle + \langle \Psi_{2p}^{\downarrow}\big(P'\big)|T^{+2}|\Psi_{2p}^{\uparrow}\big(P\big)\rangle }=0 \,.
\en
\subsection{$T^{+-}$ : up going to up matrix element}
\eq
T^{+-}=\psi_+^{\dagger}\frac{-\partial_{\perp}^2+m^2}{i\partial^+}\psi_++\frac{1}{2}(\partial^{\perp}\phi)^2+\frac{1}{2}\lambda^2\phi^2 +\text{interaction terms}\, .
\en
\eq
\langle \Psi_{2p}^{\uparrow}\big(P^+,P^{\perp}=q\big)|T_{q}^{+-}|\Psi_{2p}^{\uparrow}\big(P^+,P^{\perp}=0\big)\rangle
&=&\frac{\mathcal{F}_7^{q}(Q^2)}{I_1^q(0)},\nonumber\\
\langle \Psi_{2p}^{\uparrow}\big(P^+,P^{\perp}=q\big)|T_{b}^{+-}|\Psi_{2p}^{\uparrow}\big(P^+,P^{\perp}=0\big)\rangle
&=&\frac{\mathcal{F}_7^{b}(Q^2)}{I_1^q(0)},
\en
where
\eq
\mathcal{F}_7^{q}(Q^2)
&=&\int\frac{d^2k_{\perp}dx}{16\pi^3}  \frac{(k_{\perp}^2+m^2)}{x} 
\bigg[\psi_{+\frac{1}{2}}^{\uparrow *}(x,\vec{k'}_{\perp})\psi_{+\frac{1}{2}}^{\uparrow}(x,\vec{k}_{\perp})
+ \psi_{-\frac{1}{2}}^{\uparrow *}(x,\vec{k'}_{\perp})\psi_{-\frac{1}{2}}^{\uparrow}(x,\vec{k}_{\perp})\bigg] \, \nonumber\\
&=&\int \frac{dx}{x}  \bigg[x^{2a_1}(1-x)^{2b_1+1}\bigg\{m^2+\frac{\kappa^2(1-x)^2}{\log(1/x)}+\frac{(1-x)^2Q^2}{4}\bigg\}+\bigg(\frac{N_2}{N_1}\bigg)^2x^{2a_2-2}\nonumber\\
&&\times(1-x)^{2b_2+3}
\frac{1}{M_n^2}\bigg\{\bigg(\frac{\kappa^2}{\log(1/x)}-\frac{Q^2}{4}\bigg)m^2-\frac{Q^2(1-x)^2}{4}\bigg(\frac{\kappa^2}{\log(1/x)}+\frac{Q^2}{4}\bigg)\nonumber\\
&&+\frac{\kappa^2}{\log(1/x)}\bigg(\frac{2\kappa^2(1-x)^2}{\log(1/x)}+\frac{Q^2}{4}\bigg)\bigg\}\bigg]
\exp\bigg[-\frac{\log(1/x)}{\kappa^2}\frac{Q^2}{4}\bigg] \nonumber\\
&=& \mathbf{\mathcal{I}_{7q} }\,,
\en
\eq
\mathcal{F}_7^{b}(Q^2)
&=&\int\frac{d^2k_{\perp}dx}{16\pi^3}  \frac{1}{1-x}\bigg[\bigg(k_{\perp}-\frac{q}{2}\bigg)^2-\frac{Q^2}{4}+\lambda^2\bigg]
\bigg[\psi_{+\frac{1}{2}}^{\uparrow *}(x,\vec{k''}_{\perp})\psi_{+\frac{1}{2}}^{\uparrow}(x,\vec{k}_{\perp})\nonumber\\
&&+ \psi_{-\frac{1}{2}}^{\uparrow *}(x,\vec{k''}_{\perp})\psi_{-\frac{1}{2}}^{\uparrow}(x,\vec{k}_{\perp})\bigg] \, \nonumber\\
&=&\int dx  \bigg[x^{2a_1}(1-x)^{2b_1}\bigg\{(\lambda^2-\frac{Q^2}{4})+\frac{\kappa^2(1-x)^2}{\log(1/x)}+\frac{(1-x)^2Q^2}{4}\bigg\}+\bigg(\frac{N_2}{N_1}\bigg)^2x^{2a_2-2}\nonumber\\
&&\times(1-x)^{2b_2}
\frac{1}{M_n^2}\bigg\{\frac{2\kappa^4(1-x)^4}{(\log(1/x))^2}+\frac{\kappa^2(1-x)^2}{\log(1/x)}\big(\lambda^2-\frac{xQ^2}{2}\big)\nonumber\\
&&-\frac{x^2Q^2}{4}\bigg(\frac{Q^2}{4}(x^2-2x)+\lambda^2\bigg)\bigg\}\bigg]
\exp\bigg[-\frac{\log(1/x)}{\kappa^2(1-x)^2}\frac{x^2Q^2}{4}\bigg] \nonumber\\
&=& \mathbf{\mathcal{I}_{7b} }\,,
\en
\eq \label{eqT+-1}
\bf \langle \Psi_{2p}^{\uparrow}\big(P^+,P'^{\perp}=q\big)|T^{+-}|\Psi_{2p}^{\uparrow}\big(P^+,P^{\perp}=0\big)\rangle
=\frac{\mathcal{I}_{7q}+\mathcal{I}_{7b}}{I_1^q(0)}\,.
\en
The diquark mass $\lambda$ appears only in $\bar{C}(Q^2)$ and is taken to be $0.6$GeV.
Using the matrix elements from Eq.(\ref{tensor}),
\eq \label{eqT+-2}
&&\mathbf{\langle \Psi_{2p}^{\uparrow}\big(P^+,P'^{\perp}=q\big)|T^{+-}|\Psi_{2p}^{\uparrow}\big(P^+,P^{\perp}=0\big)\rangle} \nonumber\\
&&\mathbf{=A(Q^2)\big (2M_n^2+\frac{(q^{\perp})^2}{2}\big)-B(Q^2)\frac{(q^{\perp})^2}{2}+C(Q^2)4(q^{\perp})^2+\bar{C}(Q^2)(4M_n^2)} \,.
\en
If we ignore the $(q^{\perp})^2$ dependent term,
\eq 
\bf \langle \Psi_{2p}^{\uparrow}\big(P^+,P'^{\perp}=q\big)|T^{+-}|\Psi_{2p}^{\uparrow}\big(P^+,P^{\perp}=0\big)\rangle
=A(Q^2)(2M_n^2)+\bar{C}(Q^2)(4M_n^2)\,.
\en
So
\eq 
\bf A(Q^2)+2\bar{C}(Q^2)=\frac{1}{2M_n^2}\frac{\mathcal{I}_{7q}+\mathcal{I}_{7b}}{I_1^q(0)}\,.
\en
\subsection{$T^{+-}$ : up going to down plus down going to up matrix elements}
\eq
\langle \Psi_{2p}^{\uparrow}\big(P'\big)|T_q^{+-}|\Psi_{2p}^{\downarrow}\big(P\big)\rangle + \langle \Psi_{2p}^{\downarrow}\big(P'\big)|T_q^{+-}|\Psi_{2p}^{\uparrow}\big(P\big)\rangle &=&\frac{\mathcal{F}_8^{q}(Q^2)}{I_2^q(0)},\nonumber\\
\langle \Psi_{2p}^{\uparrow}\big(P'\big)|T_b^{+-}|\Psi_{2p}^{\downarrow}\big(P\big)\rangle + \langle \Psi_{2p}^{\downarrow}\big(P'\big)|T_b^{+-}|\Psi_{2p}^{\uparrow}\big(P\big)\rangle &=&\frac{\mathcal{F}_8^{b}(Q^2)}{I_2^q(0)},
\en
where
\eq
\mathcal{F}_8^{q}(Q^2)
&=&\int\frac{d^2k_{\perp}dx}{16\pi^3}  \frac{(k_{\perp}^2+m^2)}{x}
\bigg[\big\{\psi_{+\frac{1}{2}}^{\uparrow *}(x,\vec{k'}_{\perp})\psi_{+\frac{1}{2}}^{\downarrow}(x,\vec{k}_{\perp})
+ \psi_{-\frac{1}{2}}^{\uparrow *}(x,\vec{k'}_{\perp})\psi_{-\frac{1}{2}}^{\downarrow}(x,\vec{k}_{\perp})\big\}\nonumber\\
&&+\big\{\psi_{+\frac{1}{2}}^{\downarrow *}(x,\vec{k'}_{\perp})\psi_{+\frac{1}{2}}^{\uparrow}(x,\vec{k}_{\perp})
+ \psi_{-\frac{1}{2}}^{\downarrow *}(x,\vec{k'}_{\perp})\psi_{-\frac{1}{2}}^{\uparrow}(x,\vec{k}_{\perp})\big\}\bigg] \, \nonumber\\
&=&2 iq^2_{\perp}~ \int dx~ \frac{N_2}{N_1} \frac{1}{M_n} x^{a_1+a_2-2}(1-x)^{b_1+b_2+2}\bigg[\frac{\kappa^2(1-x)^2}{\log(1/x)}+\frac{Q^2(1-x)^2}{4}+m^2\bigg]\nonumber\\
&&\times\exp\bigg[-\frac{\log(1/x)}{\kappa^2}\frac{Q^2}{4}\bigg] \nonumber\\
&=& \bf 2 iq^2_{\perp}\mathcal{I}_{8q} \,,
\en
\eq
\mathcal{F}_8^{b}(Q^2)
&=&\int\frac{d^2k_{\perp}dx}{16\pi^3}   \frac{1}{1-x}\bigg[\bigg(k_{\perp}-\frac{q}{2}\bigg)^2-\frac{Q^2}{4}+\lambda^2\bigg]
\bigg[\big\{\psi_{+\frac{1}{2}}^{\uparrow *}(x,\vec{k''}_{\perp})\psi_{+\frac{1}{2}}^{\downarrow}(x,\vec{k}_{\perp})\nonumber\\
&&+ \psi_{-\frac{1}{2}}^{\uparrow *}(x,\vec{k''}_{\perp})\psi_{-\frac{1}{2}}^{\downarrow}(x,\vec{k}_{\perp})\big\}
+\big\{\psi_{+\frac{1}{2}}^{\downarrow *}(x,\vec{k''}_{\perp})\psi_{+\frac{1}{2}}^{\uparrow}(x,\vec{k}_{\perp})
+ \psi_{-\frac{1}{2}}^{\downarrow *}(x,\vec{k''}_{\perp})\psi_{-\frac{1}{2}}^{\uparrow}(x,\vec{k}_{\perp})\big\}\bigg] \, \nonumber\\
&=&-2 iq^2_{\perp} \int dx~ \frac{N_2}{N_1} \frac{1}{M_n} x^{a_1+a_2}(1-x)^{b_1+b_2}\bigg[\frac{\kappa^2(1-x)^2}{\log(1/x)}+\frac{Q^2(1-x)^2}{4}+(\lambda^2-\frac{Q^2}{4})\bigg]\nonumber\\
&&\times\exp\bigg[-\frac{\log(1/x)}{\kappa^2(1-x)^2}\frac{x^2Q^2}{4}\bigg] \nonumber\\
&=& \bf -2P^+ i(q^2_{\perp})\mathcal{I}_{8b} \,.
\en
\eq
\bf{\langle \Psi_{2p}^{\uparrow}\big(P'\big)|T^{+-}|\Psi_{2p}^{\downarrow}\big(P\big)\rangle + \langle \Psi_{2p}^{\downarrow}\big(P'\big)|T^{+-}|\Psi_{2p}^{\uparrow}\big(P\big)\rangle }=\frac{2(\mathcal{I}_{8q}-\mathcal{I}_{8b})}{I_2^q(0)} (iq^2_{\perp}) \,.
\en
Using the matrix elements from Eq.(\ref{tensor}),
\eq
&&\mathbf{\langle \Psi_{2p}^{\uparrow}\big(P'\big)|T^{+-}|\Psi_{2p}^{\downarrow}\big(P\big)\rangle + \langle \Psi_{2p}^{\downarrow}\big(P'\big)|T^{+-}|\Psi_{2p}^{\uparrow}\big(P\big)\rangle }\nonumber\\
&&\mathbf{=\big[A(Q^2)(2M_n)-B(Q^2)\frac{(q^{\perp})^2}{M_n}+C(Q^2)\frac{4(q^{\perp})^2}{M_n}+\bar{C}(Q^2)(4M_n)\big](-iq^2_{\perp})} \,.
\en
Ignoring the $(q^2_{\perp})^2$ dependent term,
\eq
&&\mathbf{\langle \Psi_{2p}^{\uparrow}\big(P'\big)|T^{+-}|\Psi_{2p}^{\downarrow}\big(P\big)\rangle + \langle \Psi_{2p}^{\downarrow}\big(P'\big)|T^{+-}|\Psi_{2p}^{\uparrow}\big(P\big)\rangle }\nonumber\\
&&\mathbf{=\big[A(Q^2)(2M_n)+\bar{C}(Q^2)(4M_n)\big](-iq^2_{\perp})} \,.
\en
So
\eq 
\bf A(Q^2)+2\bar{C}(Q^2)=-\frac{1}{M_n}\frac{\mathcal{I}_{8q}-\mathcal{I}_{8b}}{I_2^q(0)}\,.
\en
The interaction terms will not contribute in the $2\rightarrow2$ process. It will contribute only when we consider the higher order corrections.

\section{Matrix elements of $T^{+2}$ for nonzero skewness}\label{appa_nz}
To calculate the matrix element of Pauli-Lubanski operator, first we need to evaluate the matrix elements of $T^{+2}$ for nonzero skewness.  
For this section we use the following frame as
\eq
&&P=(P^+,P_{\perp},P^-)=\bigg(P^+,0,\frac{M^2}{P^+}\bigg) \, ,\nonumber\\
&&P'=(P'^+,P'_{\perp},P'^-)=\bigg((1-\zeta)P^+,-q_{\perp},\frac{q_{\perp}^2+M^2}{(1-\zeta)P^+}\bigg) \, ,\nonumber\\
&&q=P-P'=\bigg(\zeta P^+,q_{\perp},\frac{t+q_{\perp}^2}{\zeta P^+}\bigg) \, ,
\en
where $t=-\frac{\zeta^2M^2+q_{\perp}^2}{1-\zeta}$ and $q_{\perp}^2=Q^2$.
\subsection{$T^{+2}$ : up going to down plus down going to up matrix elements}

\eq
\langle \Psi_{2p}^{\uparrow}\big(P'\big)|T_f^{+2}|\Psi_{2p}^{\downarrow}\big(P\big)\rangle + \langle \Psi_{2p}^{\downarrow}\big(P'\big)|T_f^{+2}|\Psi_{2p}^{\uparrow}\big(P\big)\rangle &=&\frac{\mathcal{F}_9^{q}(\zeta,Q^2)}{I_2^q(0)},\nonumber\\
\langle \Psi_{2p}^{\uparrow}\big(P'\big)|T_b^{+2}|\Psi_{2p}^{\downarrow}\big(P\big)\rangle + \langle \Psi_{2p}^{\downarrow}\big(P'\big)|T_b^{+2}|\Psi_{2p}^{\uparrow}\big(P\big)\rangle &=&\frac{\mathcal{F}_9^{b}(\zeta,Q^2)}{I_2^q(0)},
\en
where
\eq
&&\mathcal{F}_9^{q}(\zeta,Q^2)\nonumber\\
&=&2P^+\int\frac{d^2k_{\perp}dx}{16\pi^3} \bigg(\frac{1-x'}{1-x}\bigg)^{1/2} \sqrt{1-\zeta} (-k_2^{\perp})
\bigg[\big\{\psi_{+\frac{1}{2}}^{\uparrow *}(x',\vec{k'}_{\perp})\psi_{+\frac{1}{2}}^{\downarrow}(x,\vec{k}_{\perp})\nonumber\\
&+& \psi_{-\frac{1}{2}}^{\uparrow *}(x',\vec{k'}_{\perp})\psi_{-\frac{1}{2}}^{\downarrow}(x,\vec{k}_{\perp})\big\}
+\big\{\psi_{+\frac{1}{2}}^{\downarrow *}(x',\vec{k'}_{\perp})\psi_{+\frac{1}{2}}^{\uparrow}(x,\vec{k}_{\perp})
+ \psi_{-\frac{1}{2}}^{\downarrow *}(x',\vec{k'}_{\perp})\psi_{-\frac{1}{2}}^{\uparrow}(x,\vec{k}_{\perp})\big\}\bigg] \, \nonumber\\
&=&-4iP^+ \frac{N_2}{N_1} \frac{1}{\kappa^2M_n} \int dx~\bigg(\frac{1-x'}{1-x}\bigg)^{1/2} \sqrt{1-\zeta} \bigg[\frac{\log x\log x'}{(1-x)(1-x')}\bigg]^{1/2} \Big\{ \big[x^{a_1}(1-x)^{b_1}\nonumber\\
&& x'^{a_2-1}(1-x')^{b_2}-x'^{a_1}(1-x')^{b_1}x^{a_2-1}(1-x)^{b_2}\big]\Big[\frac{1}{2A^2}+(q^2_{\perp})^2\frac{(\log x')^2}{4\kappa^4(1-x')^2 A_q^3}\Big]\nonumber\\
&+&x^{a_1}(1-x)^{b_1}x'^{a_2-1}(1-x')^{b_2+1}(q^2_{\perp})^2\frac{\log x'}{2\kappa^2(1-x') A_q^2}\Big\}\exp\bigg[\frac{Q^2\log x'}{2\kappa^2}\Big(\frac{\log x'}{2\kappa^2(1-x')^2 A_q}+1\Big)\bigg]\nonumber\\
&=&\bf iP^+ \mathcal{I}_{9q}^I(\zeta,Q^2) +iP^+(q^2_{\perp})^2\mathcal{I}_{9q}^{II}(\zeta,Q^2) \,,
\en
\eq
&&\mathcal{F}_9^{b}(\zeta,Q^2)\nonumber\\
&=&2P^+\int\frac{d^2k_{\perp}dx}{16\pi^3} \bigg(\frac{1-x''}{1-x}\bigg)^{1/2} \sqrt{1-\zeta}~ k_2^{\perp}
\bigg[\big\{\psi_{+\frac{1}{2}}^{\uparrow *}(x'',\vec{k''}_{\perp})\psi_{+\frac{1}{2}}^{\downarrow}(x,\vec{k}_{\perp})\nonumber\\
&+& \psi_{-\frac{1}{2}}^{\uparrow *}(x'',\vec{k''}_{\perp})\psi_{-\frac{1}{2}}^{\downarrow}(x,\vec{k}_{\perp})\big\}
+\big\{\psi_{+\frac{1}{2}}^{\downarrow *}(x'',\vec{k''}_{\perp})\psi_{+\frac{1}{2}}^{\uparrow}(x,\vec{k}_{\perp})
+ \psi_{-\frac{1}{2}}^{\downarrow *}(x'',\vec{k''}_{\perp})\psi_{-\frac{1}{2}}^{\uparrow}(x,\vec{k}_{\perp})\big\}\bigg] \, \nonumber\\
&=&4iP^+ \frac{N_2}{N_1} \frac{1}{\kappa^2M_n} \int dx~\bigg(\frac{1-x''}{1-x}\bigg)^{1/2} \sqrt{1-\zeta} \bigg[\frac{\log x\log x''}{(1-x)(1-x'')}\bigg]^{1/2} \Big\{ \big[x^{a_1}(1-x)^{b_1}\nonumber\\
&& x''^{a_2-1}(1-x'')^{b_2}-x''^{a_1}(1-x'')^{b_1}x^{a_2-1}(1-x)^{b_2}\big]\Big[\frac{1}{2A_b^2}+(q^2_{\perp})^2\frac{x''^2(\log x'')^2}{4\kappa^4(1-x'')^4 A'^3}\Big]\nonumber\\
&+&x^{a_1}(1-x)^{b_1}x''^{a_2}(1-x'')^{b_2}(q^2_{\perp})^2\frac{x''\log x''}{2\kappa^2(1-x'')^2 A_b^2}\Big\}\exp\bigg[\frac{Q^2\log x''}{2\kappa^2}\Big(\frac{x''^2\log x''}{2\kappa^2(1-x'')^4 A_b}+1\Big)\bigg]\nonumber\\
&=&\bf iP^+ \mathcal{I}_{9b}^I(\zeta,Q^2) +iP^+(q^2_{\perp})^2\mathcal{I}_{9b}^{II}(\zeta,Q^2) \,,
\en
with
\eq
A_q(x)&=&\frac{\log(1/x')}{2\kappa^2(1-x')}+\frac{\log(1/x)}{2\kappa^2(1-x)}\nonumber \, ,\\
A_b(x)&=&\frac{\log(1/x'')}{2\kappa^2(1-x'')}+\frac{\log(1/x)}{2\kappa^2(1-x)} \, ,
\en
where $x'=\frac{x-\zeta}{1-\zeta}$ and $\vec{k'}_{\perp}=\vec{k}_{\perp}-\frac{1-x}{1-\zeta}\vec{q}_{\perp}$ for the struck quark and $x''=\frac{x}{1-\zeta}$ and $\vec{k''}_{\perp}=\vec{k}_{\perp}+\frac{x}{1-\zeta}\vec{q}_{\perp}$ for the scalar diquark. So,
\eq
&&\bf{\langle \Psi_{2p}^{\uparrow}\big(P'\big)|T^{+2}|\Psi_{2p}^{\downarrow}\big(P\big)\rangle + \langle \Psi_{2p}^{\downarrow}\big(P'\big)|T^{+2}|\Psi_{2p}^{\uparrow}\big(P\big)\rangle }\nonumber\\
&=&\bf \frac{(\mathcal{I}_{9q}^{I}+\mathcal{I}_{9b}^{I})}{I_2^q(0)}(iP^+)+ \frac{(\mathcal{I}_{9q}^{II}+\mathcal{I}_{9b}^{II})}{I_2^q(0)}(iP^+(q^2_{\perp})^2)\,.\label{B5}
\en
Using the matrix elements from Eq.(\ref{tensor}):
\eq
&&\bf{\langle \Psi_{2p}^{\uparrow}\big(P'\big)|T^{+2}|\Psi_{2p}^{\downarrow}\big(P\big)\rangle + \langle \Psi_{2p}^{\downarrow}\big(P'\big)|T^{+2}|\Psi_{2p}^{\uparrow}\big(P\big)\rangle }\nonumber\\
&=&\bf\frac{1}{2}\Big[A(Q^2)+B(Q^2)\Big]\frac{\zeta(2-\zeta)}{\sqrt{1-\zeta}}M_n(iP^+) \nonumber\\
&+&\bf\Big[B(Q^2)\frac{2-\zeta}{4M_n\sqrt{1-\zeta}}+C(Q^2)\frac{\zeta}{M_n\sqrt{1-\zeta}}\Big]iP^+(q^2_{\perp})^2 \,.\label{B6}
\en
Ignoring the term $(q^2_{\perp})^2$ and comparing Eqs.(\ref{B5}) and (\ref{B6}),
\eq
\bf \frac{(\mathcal{I}_{9q}^{I}+\mathcal{I}_{9b}^{I})}{I_2^q(0)}=\frac{1}{2}\Big[A(Q^2)+B(Q^2)\Big]\frac{\zeta(2-\zeta)}{\sqrt{1-\zeta}}M_n.
\en
So
\eq
&&\bf\Big[\frac{\partial}{\partial q_-}({\langle \Psi_{2p}^{\uparrow}\big(P'\big)|T^{+2}|\Psi_{2p}^{\downarrow}\big(P\big)\rangle + \langle \Psi_{2p}^{\downarrow}\big(P'\big)|T^{+2}|\Psi_{2p}^{\uparrow}\big(P\big)\rangle })\Big]_{q=0} \,.\nonumber\\
&=&i\bf\Big[\frac{\partial}{\partial \zeta}\frac{(\mathcal{I}_{9q}^{I}+\mathcal{I}_{9b}^{I})}{I_2^q(0)}\Big]_{q=0}=iM_n[A(0)+B(0)].
\en


\end{document}